\documentclass{aa} 
\usepackage{graphicx}
\usepackage{gensymb}
\usepackage{multirow}
\usepackage{txfonts}
\usepackage{xcolor}

\usepackage[hidelinks]{hyperref}
\hypersetup{
    colorlinks=true,
    allcolors=blue,}

\begin{document}

  \title{
 Classifying merger stages with adaptive deep learning and cosmological hydrodynamical simulations
  }

  \subtitle{}

  \author{Rosa de Graaff\inst{1}, Berta Margalef-Bentabol\inst{2}, 
  Lingyu Wang\inst{1,2}\thanks{\email{l.wang@sron.nl}},   
  Antonio La Marca\inst{1, 2}, William J. Pearson\inst{3}, Vicente Rodriguez-Gomez\inst{4}, and Mike Walmsley\inst{5} 
     }

  \institute{Kapteyn Astronomical Institute, University of Groningen, Postbus 800, 9700 AV Groningen, The Netherlands 
  \email{l.wang@sron.nl}
  \and
  SRON Netherlands Institute for Space Research, Landleven 12, 9747 AD Groningen, The Netherlands
  \and
  National Centre for Nuclear Research, Pasteura 7, 02-093 Warszawa, Poland
  \and        
   Instituto de Radioastronom\'ia y Astrof\'isica, Universidad Nacional Aut\'onoma de M\'exico, Apdo. Postal 72-3, 58089 Morelia, Mexico
   \and
   Dunlap Institute for Astronomy and Astrophysics, University of Toronto, 50 St. George Street, Toronto, ON M5S 3H4, Canada
       }

  \date{Received -; accepted -}

 
 \abstract
  {}
  {Hierarchical merging of galaxies plays an important role in galaxy formation and evolution. Mergers could trigger key evolutionary phases such as starburst activities and active accretion periods onto supermassive black holes at the centres of galaxies. We aim to detect mergers and merger stages (pre- and post-mergers) across cosmic history. Our main goal is to test whether it is more beneficial to detect mergers and their merger stages simultaneously or hierarchically. In addition, we want to test the impact of merger time relative to the coalescence of  merging galaxies.}
  {First, we generated realistic mock {\it James Webb} Space Telescope ({\it JWST}) images of simulated galaxies selected from the IllustrisTNG cosmological hydrodynamical simulations. The advantage of using simulations is that we have information on both whether a galaxy is a merger and its exact merger stage (i.e. when in the past or in the future the galaxy has experienced or will experience a merging event). Then we trained deep learning (DL) models for galaxy morphology classifications in the Zoobot Python package to classify galaxies into merging/non-merging galaxies and their merger stages. We used two different set-ups, two-stage versus one-stage. In the former set-up, we first classify galaxies into mergers and non-mergers and then classify the mergers into pre-mergers and post-mergers. In the latter set-up, merger/non-merger and merger stages are classified simultaneously.  
  }
  {We found that the one-stage classification set-up moderately outperforms the two-stage set-up, offering better overall accuracy and generally better precision particularly for the non-merger class. Out of the three classes, pre-mergers can be classified with the highest precision ($\sim65\%$ vs $\sim33\%$ from a random classifier) in both set-ups, possibly due to the generally more recognisable merging features and the presence of merging companions.  More confusion is found between post-mergers and non-mergers than between these two classes and pre-mergers. The image signal-to-noise ratio (S/N) also affects the performance of the DL classifiers, but not much after a certain threshold is crossed (S/N $\sim20$ in a 0.2\arcsec aperture). In terms of merger timescale, both precision and recall of the classifiers depend strongly on merger time. Both set-ups find it more difficult to identify true mergers observed at stages that are more distant to coalescence either in the past or in the future. For pre-mergers, we recommend selecting mergers which will merge in the next 0.4 Gyrs, to achieve a good balance between precision and recall. 
  }
  {}


\titlerunning{Classifying merger stages}
\authorrunning{R. de Graaff, B. Margalef-Bentabol, L. Wang et al.}

  \maketitle
%

\section{Introduction}

In the widely-accepted hierarchical structure formation paradigm \citep{1991ApJ...379...52W, 2000MNRAS.319..168C}, galaxies collide and merge with each other throughout cosmic time, which fundamentally impact on the way galaxies form and evolve. This process can be and is still seen today in the very local Universe. For example, our own Milky Way has had several merging events in the past as revealed via stellar streams and is heading towards a collision course with the neighbouring Andromeda galaxy in the future (review by \citealt{2020ARA&A..58..205H} and references therein). Over the past several decades, there have been numerous studies on measuring the merger history of galaxies all the way from the present-day Universe to the epoch of reionisation \citep{2003AJ....126.1183C, 2011ApJ...742..103L, 2019ApJ...876..110D, 2020ApJ...895..115F, 2022ApJ...940..168C} as well as deciphering the role of merging in galaxy mass assembly, morphological transformation, triggering of accretion episodes onto supermassive black holes (SMBH) and star-formation activity \citep{2008AJ....135.1877E, 2011MNRAS.412..591P, 2014MNRAS.441.1297S, 2018PASJ...70S..37G, 2019A&A...631A..51P, 2019MNRAS.487.2491E, 2020A&A...637A..94G, 2022A&A...661A..52P, 2023OJAp....6E..34V, lamarcaDustPowerUnravelling2024}. 

However, despite the huge progress already made, we still do not understand the relative importance of mergers and secular evolution particularly in the more distant Universe or if there are definitive links (and how they evolve) between mergers and key galaxy evolution phases such as starbursts and active galactic nuclei (AGN). One of the main challenges that studies involving galaxy mergers face is merger detection. Various methods have been used in the past to identify merging galaxies, e.g., counting galaxy pairs in photometric or spectroscopic data \citep{2012A&A...539A..45L, 2015MNRAS.454.1742K, 2024arXiv240709472D}, and detecting morphologically disturbed galaxies using non-parametric statistics, such as the concentration-asymmetry-smoothness (CAS) parameters, M20 (the second-order moment of the brightest 20\% of the light) and the Gini coefficient \citep{2000ApJ...529..886C, 2003ApJS..147....1C, 2004AJ....128..163L, 2008ApJ...672..177L},  or visual inspection, such as expert classification labels \citep{2015ApJS..221...11K} or volunteer labels from the Galaxy Zoo (GZ) citizen science project \citep{2008MNRAS.389.1179L, 2010MNRAS.401.1043D, 2011MNRAS.410..166L}. Different methods have different pros and cons which were extensively discussed in many previous works \citep{2015ApJS..221....8H, 2019MNRAS.486.3702S, 2020MNRAS.492.2075B, 2024A&A...687A..24M}. 
According to hydrodynamical simulations, starburst, quenching of star formation, and accretion onto SMBHs take place at different stages \citep{2008ApJS..175..356H, 2018MNRAS.478.3056B, 2019MNRAS.490.2139R, 2023MNRAS.520.4463L, 2023MNRAS.519.2119Q, 2024MNRAS.528.5864B}.
Therefore, beyond a binary classification of merging and non-merging galaxies, it is also  important to place merging galaxies along their merging sequences which typically last of the order a couple of  billion years.  However, so far, our knowledge of merger stages beyond a basic pre-merger and post-merger separation (i.e., merging galaxies caught before and after coalescence) is extremely limited.

With the advent of deep learning (DL; \citealt{Goodfellow-et-al-2016}) techniques which excel in computer vision tasks and increasingly more realistic cosmological hydro-dynamical simulations \citep{2014MNRAS.444.1518V, 2014MNRAS.444.1453D, 2015MNRAS.446..521S, 2019ComAC...6....2N}, some of the fundamental limitations present in previous studies can be better addressed or even overcome. Large samples of realistic merging galaxies at different stages along their merging sequences (via information on galaxy merger tree histories) are now available from cosmological simulations, which can be used to train or test merger detection algorithms with reliable truth labels. This means we are no longer limited to classification labels derived from visual inspection which is subjective, (often) unreliable, incomplete, and time-consuming to obtain. For the first time, we can quantitatively assess  the levels of completeness and reliability for any merger detection algorithms, which are essential to  statistical studies such as measurements of galaxy merger fractions and merger rates. At the same time, there has been an explosion over the past decade in applying DL techniques, in particular convolutional neural networks (CNNs; \citealt{726791, article}), to astronomical questions (such as galaxy morphological classifications and detections of strongly lensed galaxies) and many of them have shown great success \citep{2015MNRAS.450.1441D, 2015ApJS..221....8H, 2017MNRAS.472.1129P, 2018A&A...611A...2S, 2020ApJ...899...30L}. For a review, please refer to \citet{2023PASA...40....1H}.

The first example of applying DL CNNs in galaxy mergers detection was presented in \citet{2018MNRAS.479..415A}, which was trained on the imagenet data set \citep{5206848} and visual classification labels from Galaxy Zoo.
The first deep CNNs trained on cosmological hydrodynamic simulations can be found in \citet{2019A&A...626A..49P}. Since then, there has been a plethora of studies exploiting the strengths of combining DL models and cosmological simulations \citep{2019MNRAS.490.5390B, 2020A&A...644A..87W, 2020ApJ...895..115F, 2020A&C....3200390C, 2021MNRAS.506..677C, 2021MNRAS.504..372B}. Through these studies we have learned several important lessons, e.g., the importance of the training data, the need to include full observational effects, and the impact of domain adaptation and transfer learning (e.g. from simulations to observations). \citet{2024A&A...687A..24M} presented the first galaxy merger challenge in which leading machine learning (ML) and DL-based merger detection algorithms are bench marked using the same datasets constructed from two different cosmological simulations, IllustrisTNG \citep{2019ComAC...6....2N} and Horizon-AGN \citep{2014MNRAS.444.1453D}, and observations from the Hyper Suprime-Cam Subaru Strategic Program (HSC-SSP) survey \citep{2018PASJ...70S...4A}. Out of the methods tested, Zoobot \citep{2023JOSS....8.5312W}, which is a Python package implementation of adaptive DL models pre-trained on GZ visual labels, seem to offer the best overall performance particularly in terms of precision of the merger class. In terms of identification of merger stages using DL techniques, there has been relatively few studies. A multi-step classification framework was adopted in \citet{2024MNRAS.533.2547F} which identified mergers and non-mergers first and then further divided the classified mergers into interacting pairs (i.e., pre-mergers) and post-mergers in a subsequent step. They demonstrated high success rate when applying this method to galaxies from redshift $z=0$ to 0.3. \citet{2024A&A...687A..45P} investigated the possibility of recovering merger times before or after merging event for galaxies selected from the IllustrisTNG simulations at redshifts $0.07\leq z\leq 0.15$, using both ML and DL methods. They found that the best-performing method is able to recover merger times with a median uncertainty of around 200 Myr, which is similar to the time resolution of the snapshots in the TNG simulations. 

In this paper, we aim to develop a method to detect and classify galaxy merging events into pre-mergers, post-mergers, and non-mergers, using DL CNN models as implemented and pre-trained in the Zoobot package. Specifically, our primary goal is to test whether it is more beneficial to classify merging and non-merging galaxies first and then further divide mergers into different merger stages, or whether it is better to classify non-mergers and merger stages (i.e. pre- and post-mergers) at the same time. Our training data come from realistic mock images of simulated galaxies over a wide redshift range (up to $z = 3$) selected from the IllustrisTNG cosmological hydrodynamical simulations, mimicking the {\it James Webb} Space Telescope ({\it JWST}; \citealt{2006SSRv..123..485G}) Near Infrared Camera (NIRcam; \citealt{2023PASP..135b8001R}) observations in the COSMOS field.  
Two classification set-ups are trained and tested. One performs a simultaneous three-class classification and is referred to as the one-stage set-up. The other performs the classification of mergers/non-mergers and merger stages in a hierarchical fashion (similar to the multi-step framework in \citealt{2024MNRAS.533.2547F}) and is referred to as the two-stage set-up. In future studies, we will apply the best-performing set-up based on this paper to investigate whether specific galaxy evolution phases (such as starbursts, dust-obscured and unobscured AGN) preferentially occur in the pre- or post-merger stage, which can then be compared with predictions from simulations of galaxy evolution.

The paper is organised as follows. In Sect. \ref{sec: data}, we describe the selection of the merging and non-merging galaxy samples from the IllustrisTNG simulations and the steps taken to create from the synthetic images realistic mock {\it JWST} NIRCam images. In Sect. \ref{sec: method}, we briefly introduce Zoobot and the DL model we use in this study to detect mergers and non-mergers, and to further classify mergers into different merger stages. The metrics we use to quantify and compare the performance of the models are also explained. In Sect. \ref{sect: results}, we show the results from both classification set-ups. First, an overall analysis including dependence on signal-to-noise ratio (S/N) and redshift is presented. Then, we focus the analysis on the dependence of the model performance on merger time and subsequently we re-train our models by adopting new definitions of merger stages. Finally, we present our main conclusions and future directions in Sect. \ref{sect: conclusion}.

\begin{figure*}
    \centering
    \setlength{\tabcolsep}{1pt}
    \begin{tabular}{ccccccc}
        -0.8 Gyr & -0.6 Gyr & -0.4 Gyr & -0.2 Gyr & 0 Gyr & 0.2 Gyr & 0.4 Gyr \\
        \includegraphics[width=0.137\textwidth, trim={3.4cm 1.4cm 3.5cm 1.3cm}, clip]{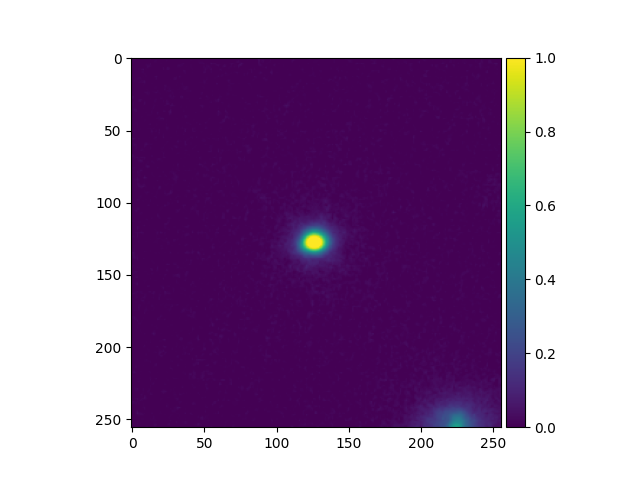} & \includegraphics[width=0.137\textwidth, trim={3.4cm 1.4cm 3.5cm 1.3cm}, clip]{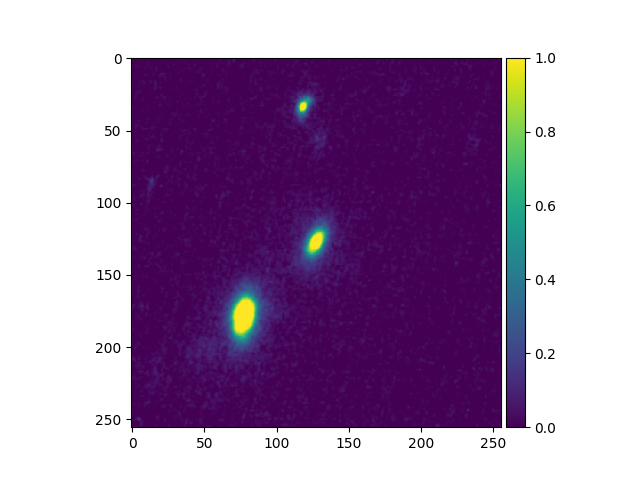} & \includegraphics[width=0.137\textwidth, trim={3.4cm 1.4cm 3.5cm 1.3cm}, clip]{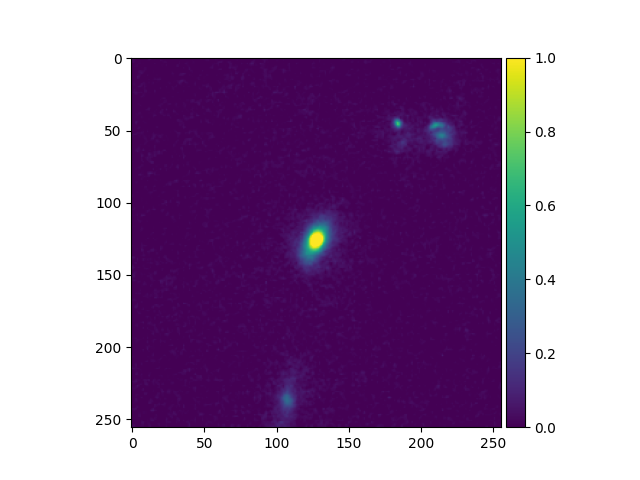} & \includegraphics[width=0.137\textwidth, trim={3.4cm 1.4cm 3.5cm 1.3cm}, clip]{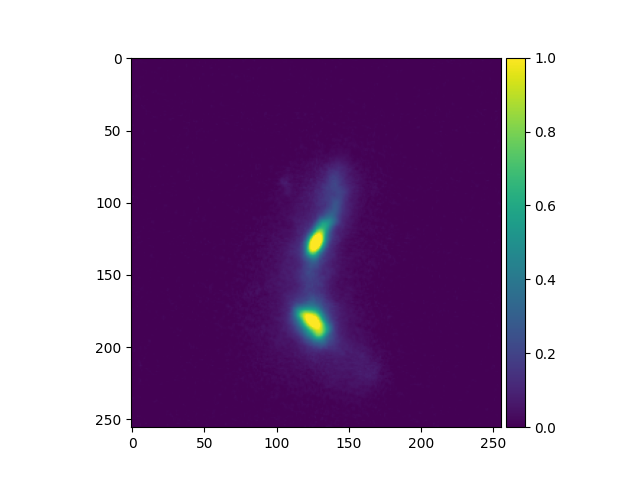} & \includegraphics[width=0.137\textwidth, trim={3.4cm 1.4cm 3.5cm 1.3cm}, clip]{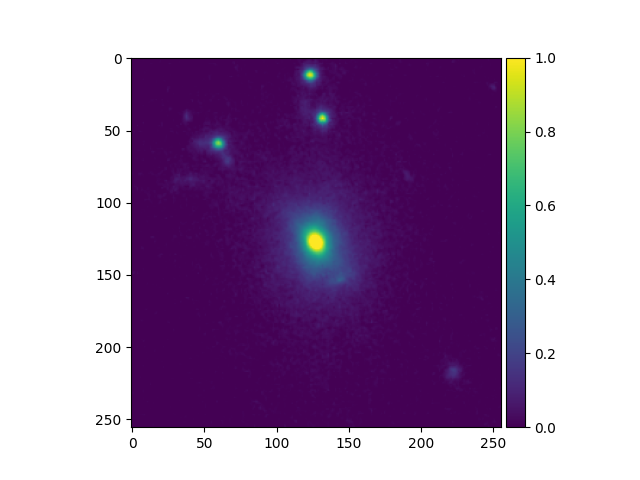} & \includegraphics[width=0.137\textwidth, trim={3.4cm 1.4cm 3.5cm 1.3cm}, clip]{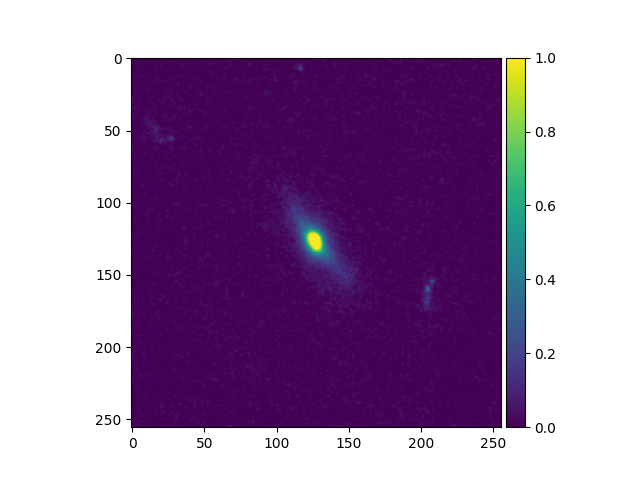} & \includegraphics[width=0.137\textwidth, trim={3.4cm 1.4cm 3.5cm 1.3cm}, clip]{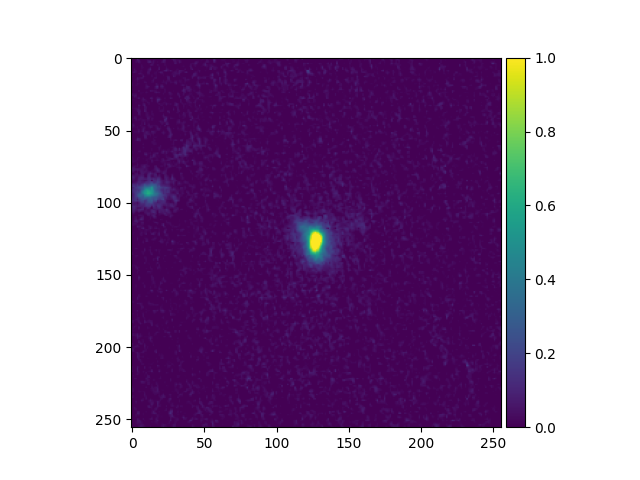}\\
        \includegraphics[width=0.137\textwidth, trim={3.4cm 1.4cm 3.5cm 1.3cm}, clip]{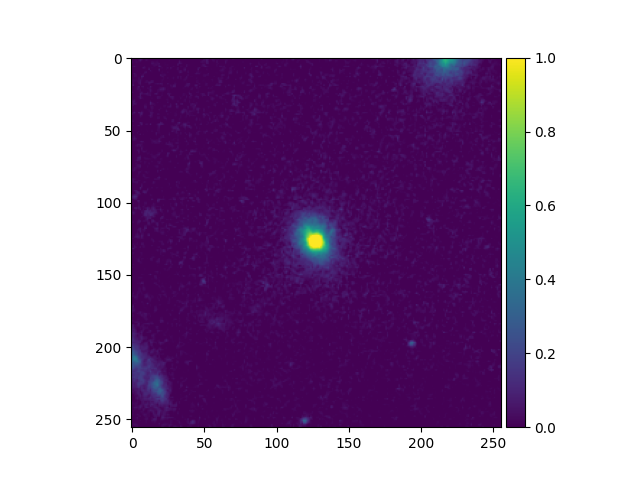} &
        \includegraphics[width=0.137\textwidth, trim={3.4cm 1.4cm 3.5cm 1.3cm}, clip]{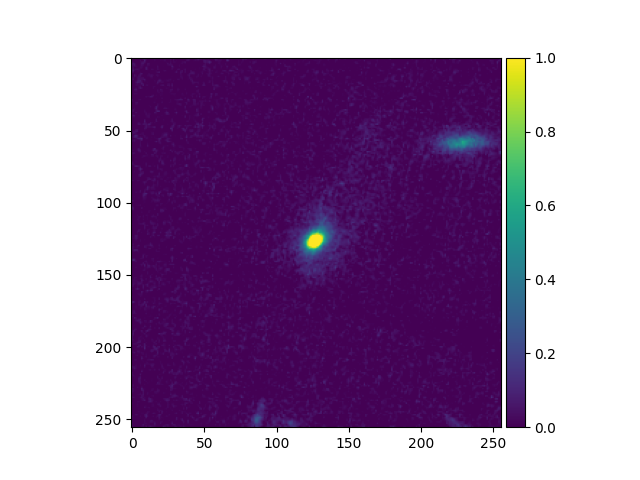} & \includegraphics[width=0.137\textwidth, trim={3.4cm 1.4cm 3.5cm 1.3cm}, clip]{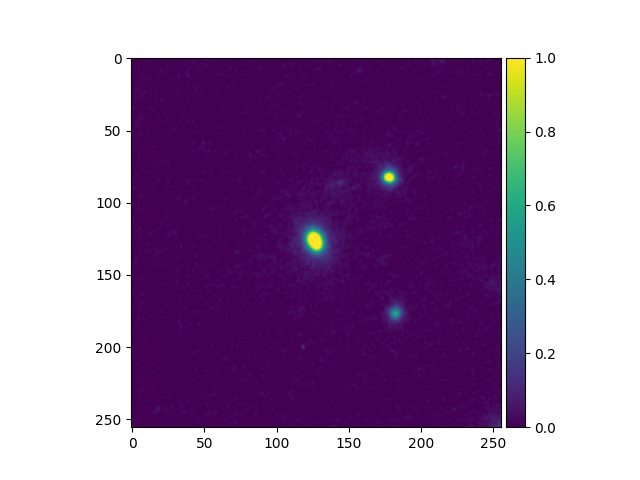} & \includegraphics[width=0.137\textwidth, trim={3.4cm 1.4cm 3.5cm 1.3cm}, clip]{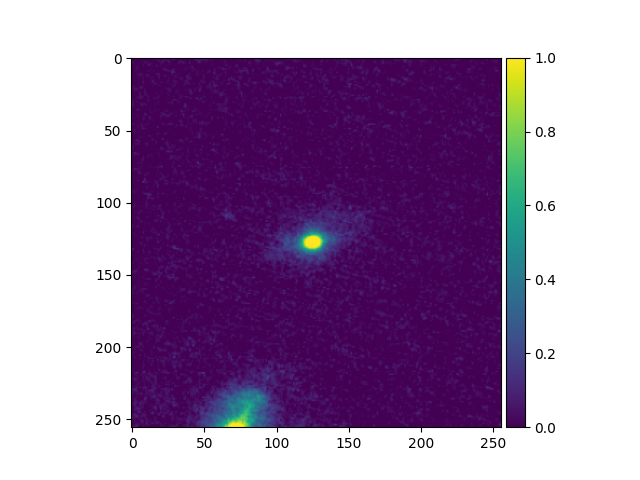} & \includegraphics[width=0.137\textwidth, trim={3.4cm 1.4cm 3.5cm 1.3cm}, clip]{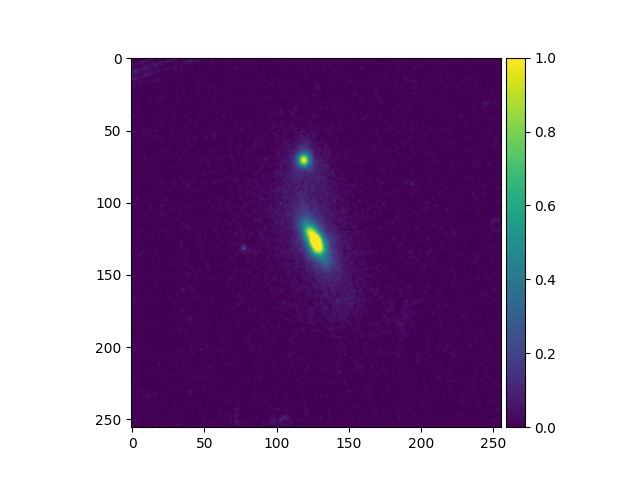} & \includegraphics[width=0.137\textwidth, trim={3.4cm 1.4cm 3.5cm 1.3cm}, clip]{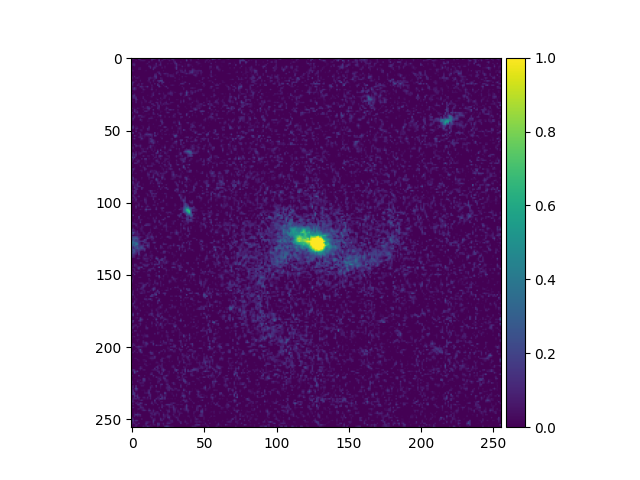} & \includegraphics[width=0.137\textwidth, trim={3.4cm 1.4cm 3.5cm 1.3cm}, clip]{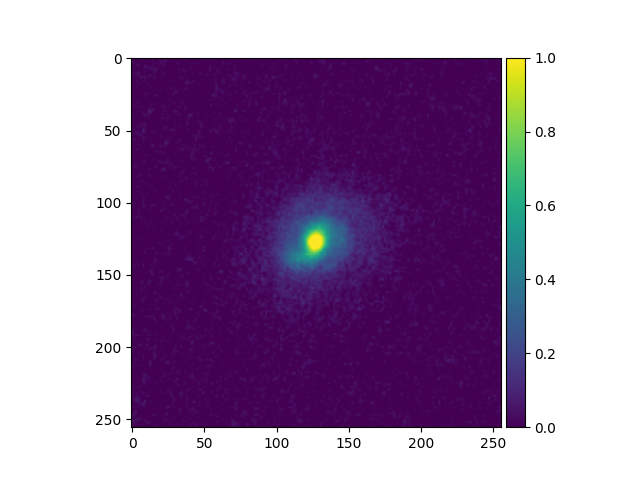}\\
        \includegraphics[width=0.137\textwidth, trim={3.4cm 1.4cm 3.5cm 1.3cm}, clip]{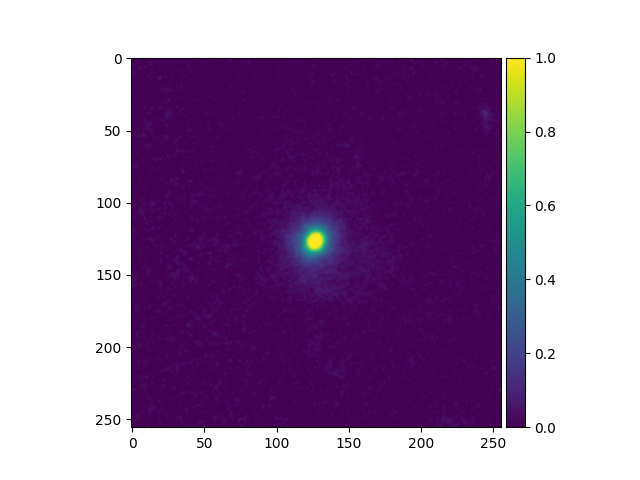} &
        \includegraphics[width=0.137\textwidth, trim={3.4cm 1.4cm 3.5cm 1.3cm}, clip]{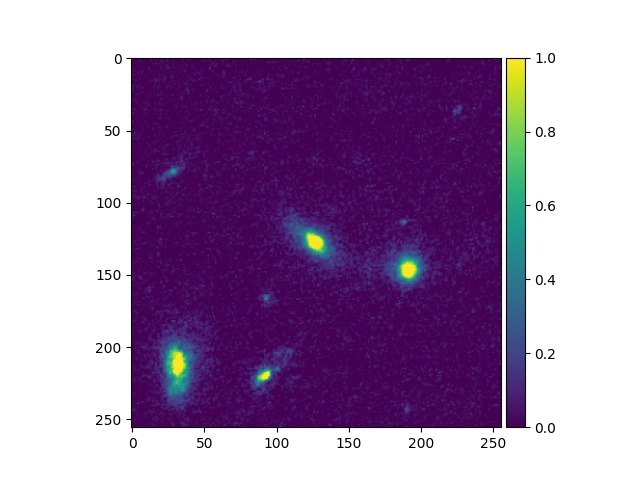} & \includegraphics[width=0.137\textwidth, trim={3.4cm 1.4cm 3.5cm 1.3cm}, clip]{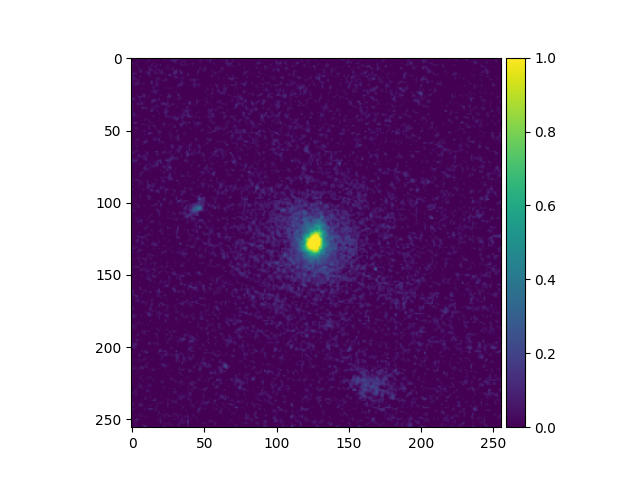} & \includegraphics[width=0.137\textwidth, trim={3.4cm 1.4cm 3.5cm 1.3cm}, clip]{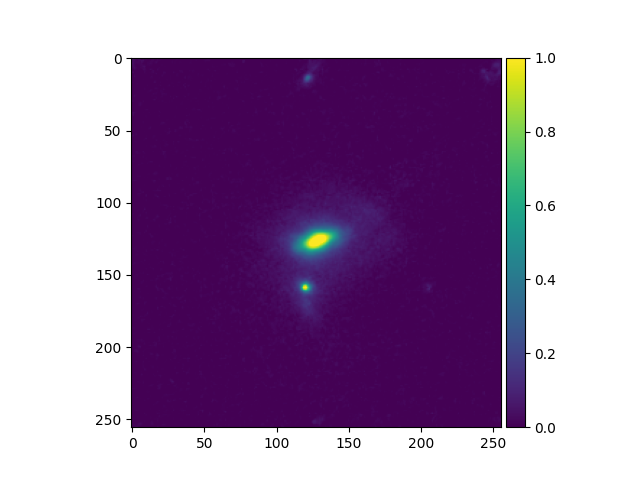} & \includegraphics[width=0.137\textwidth, trim={3.4cm 1.4cm 3.5cm 1.3cm}, clip]{ongpost_001berta.png} & \includegraphics[width=0.137\textwidth, trim={3.4cm 1.4cm 3.5cm 1.3cm}, clip]{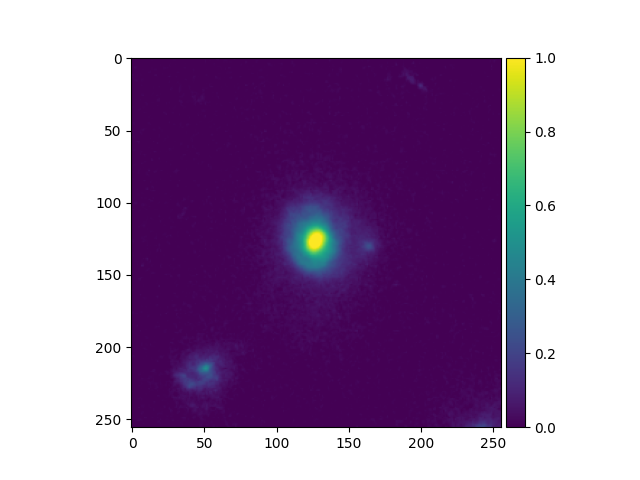} & \includegraphics[width=0.137\textwidth, trim={3.4cm 1.4cm 3.5cm 1.3cm}, clip]{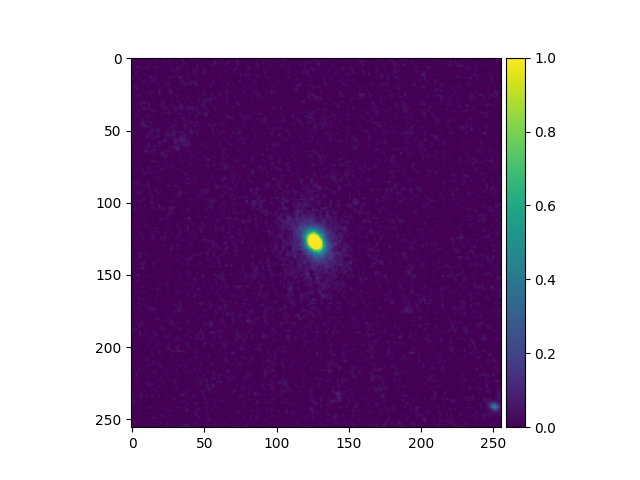}\\
        \includegraphics[width=0.137\textwidth, trim={3.4cm 1.4cm 3.5cm 1.3cm}, clip]{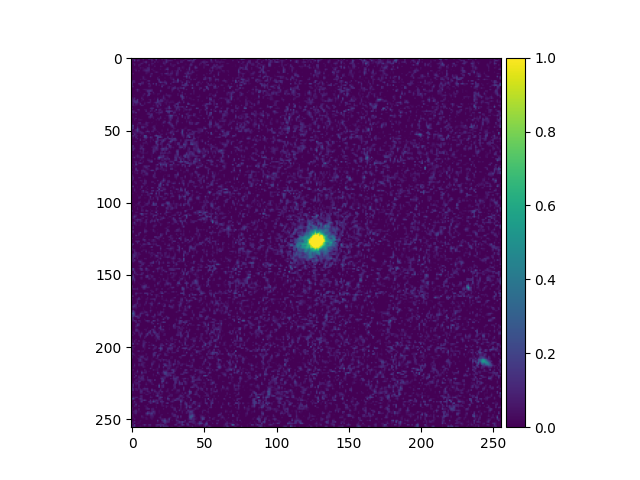} &
        \includegraphics[width=0.137\textwidth, trim={3.4cm 1.4cm 3.5cm 1.3cm}, clip]{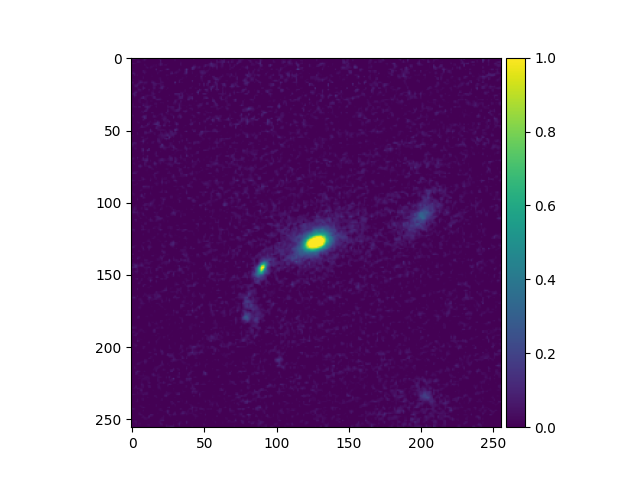} & \includegraphics[width=0.137\textwidth, trim={3.4cm 1.4cm 3.5cm 1.3cm}, clip]{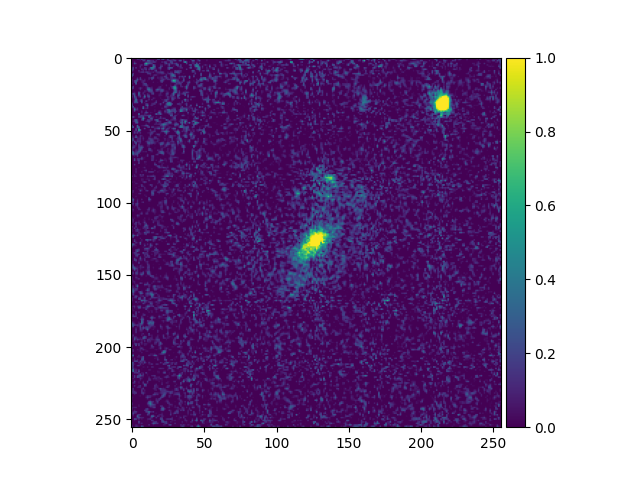} & \includegraphics[width=0.137\textwidth, trim={3.4cm 1.4cm 3.5cm 1.3cm}, clip]{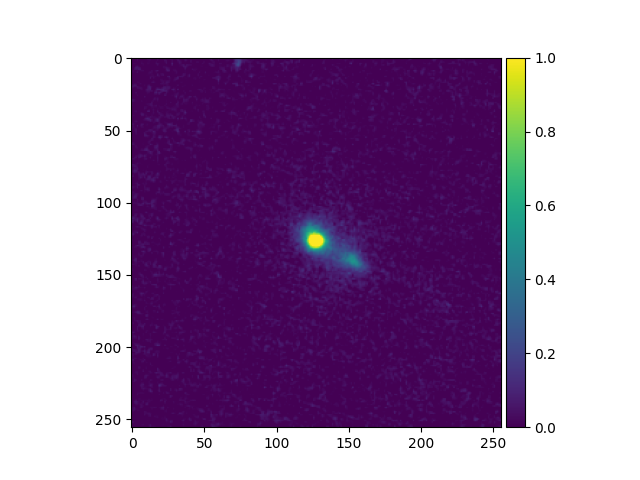} & \includegraphics[width=0.137\textwidth, trim={3.4cm 1.4cm 3.5cm 1.3cm}, clip]{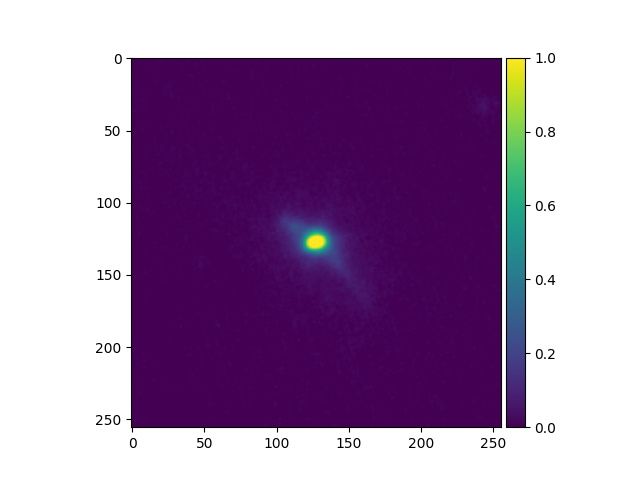} & \includegraphics[width=0.137\textwidth, trim={3.4cm 1.4cm 3.5cm 1.3cm}, clip]{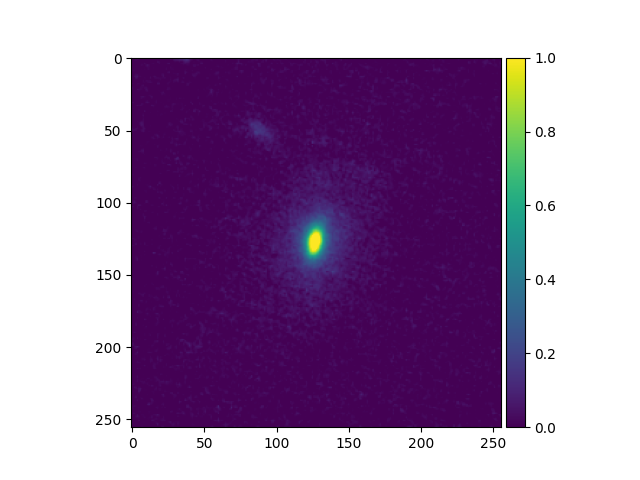} & \includegraphics[width=0.137\textwidth, trim={3.4cm 1.4cm 3.5cm 1.3cm}, clip]{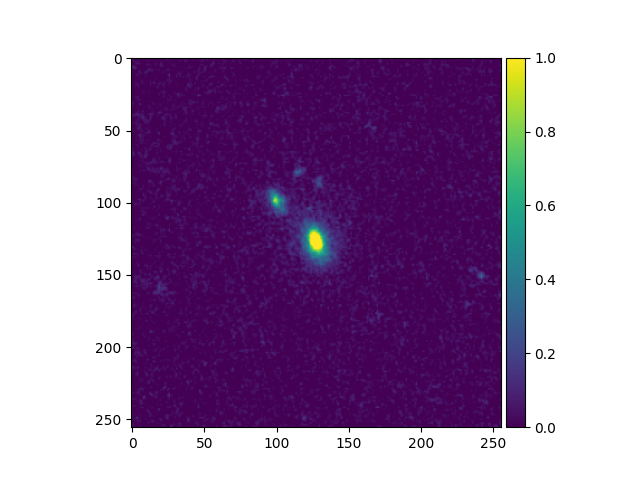}\\
        \includegraphics[width=0.137\textwidth, trim={3.4cm 1.4cm 3.5cm 1.3cm}, clip]{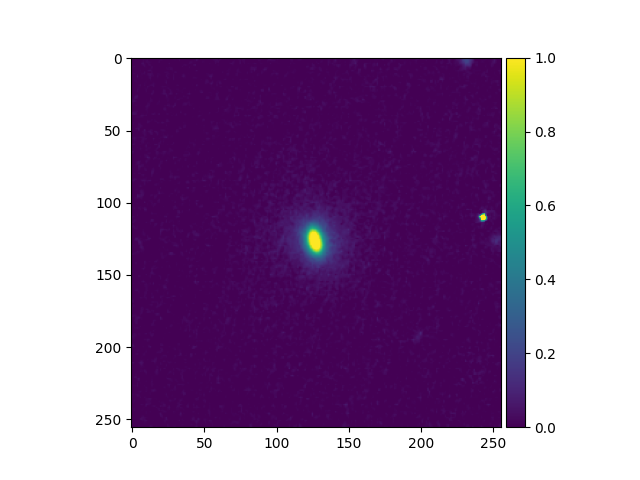} &
        \includegraphics[width=0.137\textwidth, trim={3.4cm 1.4cm 3.5cm 1.3cm}, clip]{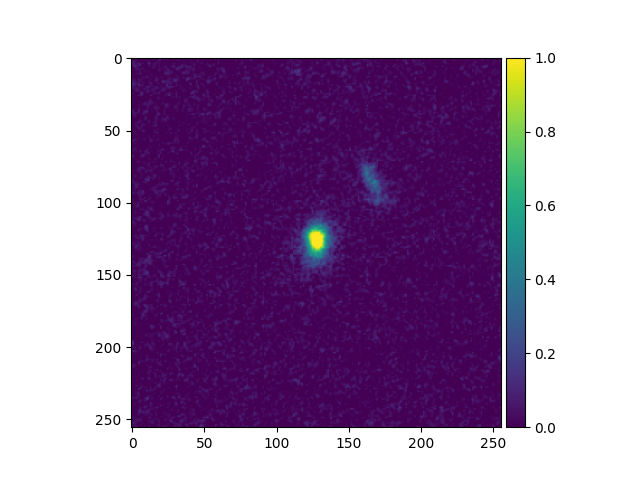} & \includegraphics[width=0.137\textwidth, trim={3.4cm 1.4cm 3.5cm 1.3cm}, clip]{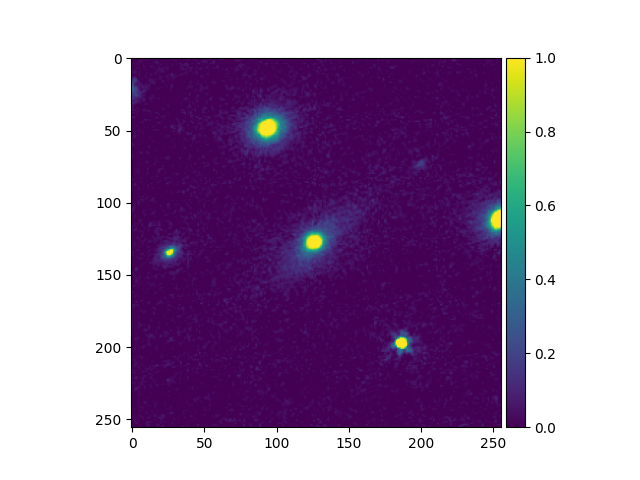} & \includegraphics[width=0.137\textwidth, trim={3.4cm 1.4cm 3.5cm 1.3cm}, clip]{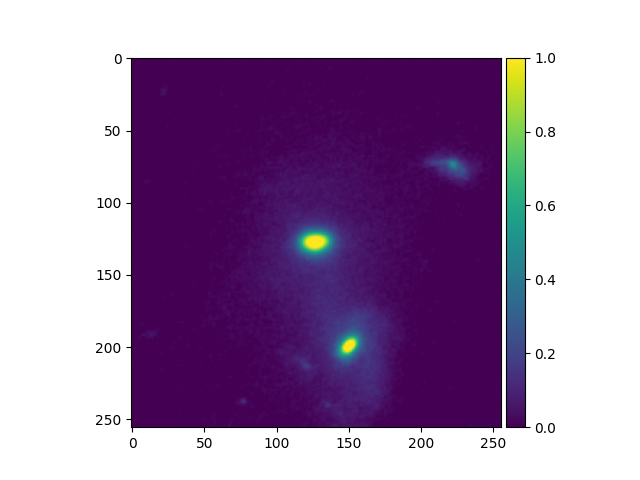} & \includegraphics[width=0.137\textwidth, trim={3.4cm 1.4cm 3.5cm 1.3cm}, clip]{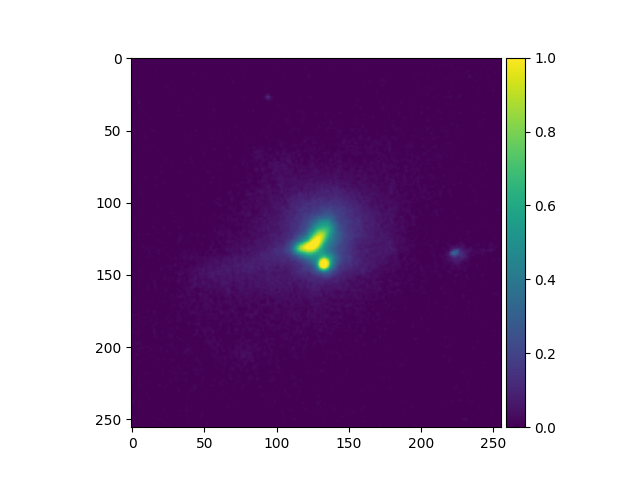} & \includegraphics[width=0.137\textwidth, trim={3.4cm 1.4cm 3.5cm 1.3cm}, clip]{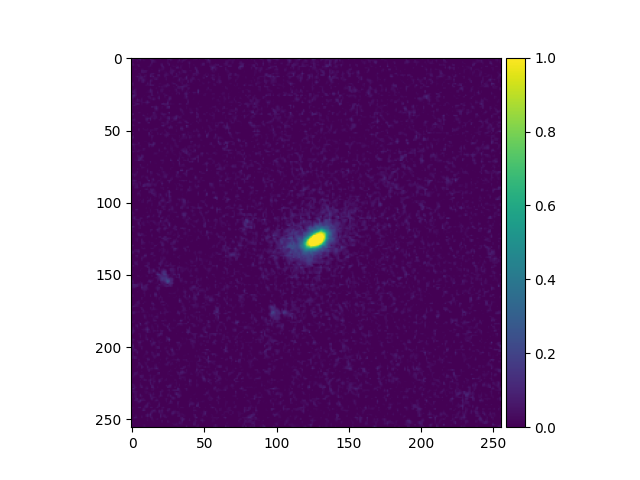} & \includegraphics[width=0.137\textwidth, trim={3.4cm 1.4cm 3.5cm 1.3cm}, clip]{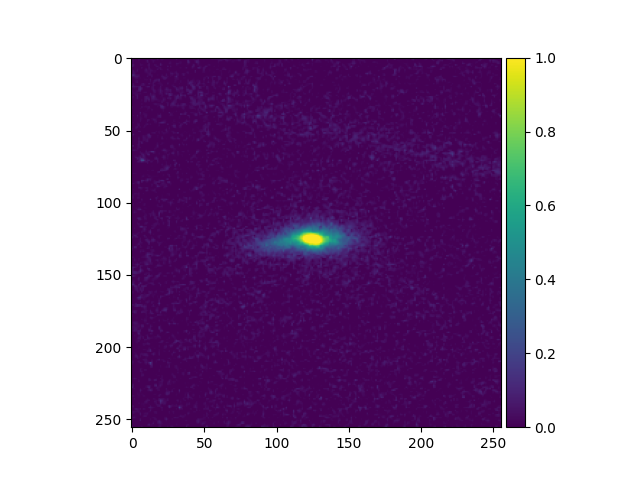}\\
    \end{tabular}
    \caption{Five examples of statistical merger sequences selected from the IllustrisTNG simulation. The exact merger time $t_{\rm merger}$ has been rounded to the nearest 0.1 Gyr. According to our default definition, pre-mergers are those with $t_{\rm merger}$ between -0.8 and -0.1 Gyr, ongoing mergers are between -0.1 and 0.1 Gyr, and post-mergers are between 0.1 and 0.3 Gyr. Galaxies which do not satisfy these conditions are defined as non-mergers. The cut-outs are 256 $\times$ 256 pixels, with a pixel resolution of $\sim0.03$\arcsec / pix. Images are scaled using the aggressive arcsinh scaling and then normalised.}
    \label{fig:merger-sequence}
\end{figure*}

\section{Data}
\label{sec: data}

Our aim is to use DL models to detect mergers in galaxy images and to  determine their merging stages. For training we need a sufficient number of images for the various classes. In this section, we first describe how samples of mergers at different merging stages and non-mergers are selected from the IllustrisTNG project. Then we explain how realistic mock {\it JWST} images mimicking various observational effects are created.

\subsection{IllustrisTNG}

\begin{figure*}
\includegraphics[width=0.33\textwidth, trim={3.4cm 1.4cm 3.5cm 1.3cm}, clip]{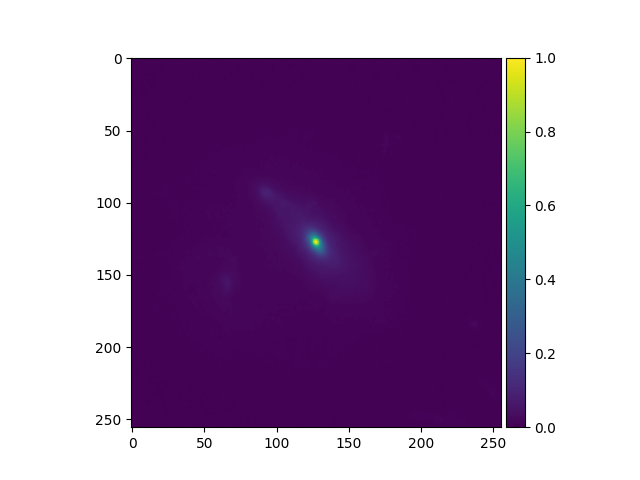}
\includegraphics[width=0.33\textwidth, trim={3.4cm 1.4cm 3.5cm 1.3cm}, clip]{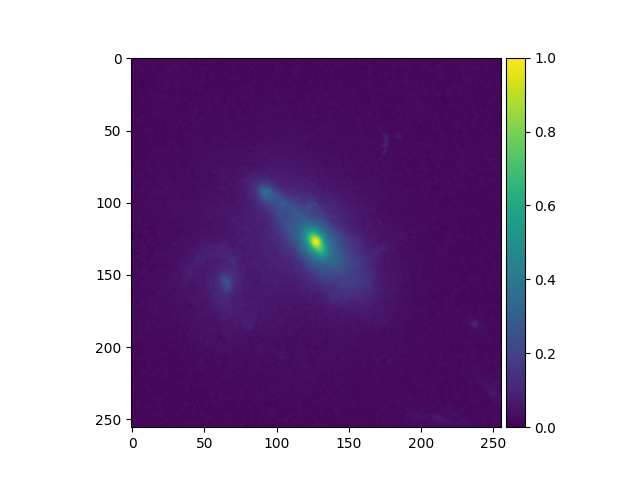}
\includegraphics[width=0.33\textwidth, trim={3.4cm 1.4cm 3.5cm 1.3cm}, clip]{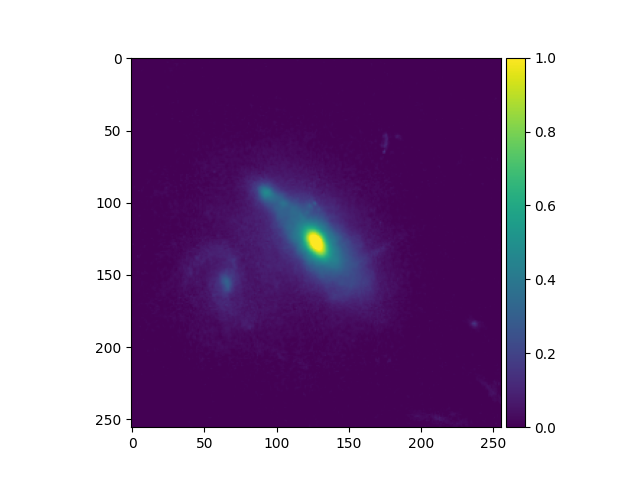}
\caption{Illustration of various scalings (left: normalised image without applying any scaling; centre: arcsinh scaling; right: aggressive arcsinh scaling) applied to the mock {\it JWST}/NIRCam F150W images. The cut-outs are 256 $\times$ 256 pixels, with a pixel resolution of $\sim0.03$\arcsec / pix. Images are normalised after each scaling.
}
\label{fig:scalings}
\end{figure*}

The IllustrisTNG project contains three cosmological hydro-dynamical simulations of galaxy formation and evolution (\citealp{Nelson_2017}; \citealp{2018MNRAS.480.5113M}; \citealp{10.1093/mnras/stx3304}; \citealp{2018MNRAS.477.1206N}; \citealp{10.1093/mnras/stx3112}). The three runs, TNG50, TNG100 and TNG300, differ in volume and resolution, with co-moving length sizes of $50, 100$ and $300$ Mpc $h^{-1}$, respectively. 
The initial conditions for the simulations are consistent with the results of the \citealp{2016A&A...594A..13P}. 
Various physical processes, such as radiative cooling, stellar feedback and evolution of stellar populations, and feedback from central SMBHs are included. 
Many studies have shown that the TNG simulations can reasonably reproduce a wide range of observed statistical galaxy properties, such as the galaxy stellar mass functions, the cosmic star-formation rate density and the galaxy star formation main sequence \citep{10.1093/mnras/stx3112, 2019MNRAS.485.4817D}. For further details, we refer the reader to \cite{nelson2021illustristng} and references therein.


For our study, we are interested in the redshift range $0.5 \leq z \leq 3.0$, corresponding to snapshot numbers $67 - 25$. The time step between each snapshot is on average $\sim$154\ \text{Myr} over this redshift interval.  We selected galaxies with stellar mass $M_{\ast} > 10^9 M_{\odot}$ from TNG100, which has a dark matter (DM) particle resolution of $M_{\mathrm{DM\, res}} = 7.5 \times10^6 M_{\odot}$ and a baryonic particle resolution of $M_{\mathrm{baryon\,res}}=1.4\times10^6$ M$_{\odot}$. For each galaxy, we made synthetic images with a physical size of $50 \times 50$ kpc in the \textit{JWST}/NIRCam F150W filter in three different projections. The process of generating the synthetic images is detailed in \citet{2024A&A...687A..24M}. Briefly, the contribution from all stellar particles around the main galaxy, with their respective spectral energy distributions based on the \citet{2003MNRAS.344.1000B} stellar population synthesis models, are summed and passed through the F150W filter to make a 2D projected map. We do not perform a full radiative transfer treatment to account for dust. However, currently most simulations do not explicitly trace the production and destruction processes of dust. Consequently, to model the effect of dust, many assumptions (e.g., on dust composition and distribution) have to be made. The validity of these assumptions remain to be tested \citep{2021MNRAS.501.4359Z}.

A complete merger history is available for each galaxy \citep{Rodriguez_Gomez_2015}, which is constructed by tracking the baryonic content of subhalos. 
Using these trees, a merger event is identified when a galaxy has more than one direct progenitor. We only considered events for which the stellar mass ratios between the progenitors are larger than $1/4$, i.e., the so-called major mergers. The stellar mass of both progenitors is taken at the moment when the secondary galaxy reaches its maximum stellar mass. 
We can further divide the mergers into three sub-classes, i.e., pre-, ongoing- and post-mergers, based on their merger times  $t_{\rm merger}$ relative to the actual merging event (i.e., coalescence of two merging galaxies into a single system). We use negative values for $t_{\rm merger}$ to denote merging events which will take place in the future and positive values for $t_{\rm merger}$ to denote merging events which happened in the past. In our default definition, pre-mergers correspond to galaxies which will have a merger event in the next $0.8$ to $0.1$ Gyrs. Therefore, in our notation, pre-mergers have $t_{\rm merger} = -0.8$ Gyr to $t_{\rm merger} = -0.1$ Gyr. In view of the average time step between each snapshots in the simulations (around 150 Myr in the redshift range we are interested in), ongoing mergers are defined as galaxies which are found to be within $0.1$ Gyrs before or after coalescence ($t_{\rm merger} = -0.1$ to 0.1 Gyr). Finally, post-mergers are defined as galaxies which have had a merger event in the previous 0.1 to 0.3 Gyrs ($t_{\rm merger} = 0.1$ to 0.3 Gyr). However, later investigations revealed poor performance of the classifiers in identifying the class of ongoing mergers due to the much smaller sample size. Therefore, we combined the ongoing and post-mergers together and from now on simply refer to this combined class as post-mergers.

In Fig. \ref{fig:merger-sequence}, we show example mock {\it JWST} NIRcam F150W images of mergers and their associated merger times (rounded to 100 Myr). Details of how these mock {\it JWST} images are generated are presented below in Sect. \ref{sect: mock}. Galaxies at different merger stages form statistical merger sequences (as opposed to merger sequences formed by following the same galaxy over its evolutionary history).  
While it is possible to follow the actual evolutionary history of a galaxy in the simulations, in real observations we can only ever catch galaxies at a single specific point along their 
evolutionary trajectories. By eye, it is clear that identification of mergers and their merger stages in these mock {\it JWST} images can be challenging. Some pre-mergers are relatively easy to identify as they show not only the presence of merging companion, but also clear interaction and tidal features. But others are difficult to distinguish from non-mergers which have a nearby galaxy by chance. The same is true for post-mergers as only some still show relatively obvious signs of merging (double nuclei, shells, tidal tails, etc.).

\subsection{Realistic mock {\it JWST} images}
\label{sect: mock}

In order to generate realistic mock {\it JWST} NIRcam F150W images, various observational effects such as spatial resolution, S/N, and background need to be included. The images from the simulation were produced with the same pixel resolution as the real {\it JWST} observations (0.03 \arcsec/pixel) and the same units of MJy/sr, and have a physical size of $50\times50$ kpc. First, we convolved each image with the observed point spread function (PSF) in the F150W filter. Then, we added shot noise as Poisson noise, to account for the statistical temporal variation of a source's photon emissions. Specifically, we used a randomly chosen global PSF model from the 80 models in total derived by \cite{zhuang2023agns}. 

To make the images more realistic, we added real background from the {\it JWST} imaging data released by the COSMOS-Web programme \citep{casey2023cosmosweb} and reduced by \cite{zhuang2023agns}. COSMOS-Web covers the central area (0.54 $deg^2$) of the COSMOS field \citep{2007ApJS..172....1S} in four NIRCam imaging filters (F115W, F150W, F277W and F444W). We made image cutouts using the observed COSMOS-Web imaging data, centred on positions which do not contain artefacts or bright sources. 
To generate these sky cutouts, we adopted the methodology outlined in \citet{2024A&A...687A..24M}. First, we compiled a list of sources to exclude, starting from the COSMOS2020 catalogue \citep{2022ApJS..258...11W}. We used the Farmer version of COSMOS2020, applying the condition that the flag \textsc{flag\_combined} $=0$, as recommended by the COSMOS team, to avoid areas contaminated by bright stars or large artefacts. Second, we filtered sources based on the lp\_type parameter, selecting those flagged as 0 or 2 (corresponding to galaxy or X-ray source). Third, we restricted our selection to sources at $z<3$ in the COSMOS-Web field, allowing distant galaxies to be included in the cutouts. Finally, we randomly generated sky coordinates, ensuring no catalogued sources (in the list to avoid) were present within a 6.5\arcsec radius. This radius was determined based on the estimated source density in the region where the cutouts were generated. 
As a result, these cutouts may still include background galaxies and faint sources. As the last step, we resized all images to $256\times 256$ pixels. This resulted in galaxies at different redshifts having slightly different pixel scales, from 0.024\arcsec/pixel at the highest redshift to 0.032 \arcsec/pixel at the lowest redshift.

We also applied image scaling to make the low surface brightness details of the galaxies more apparent. First, we considered an arcsinh scaling which reduces the dynamical range drastically, but also preserves the structures in the brighter regions  \citep{Lupton_2004}. 
We then used a more aggressive arcsinh scaling which decreased the background even further, and removed any potential outlier pixels, following the approach of \cite{2019MNRAS.490.5390B}. This was done by setting every pixel value below the median of the entire image to the median value. On top of that, pixel values above the $90th$ percentile of the pixel intensities in the centre of the image ($80\times80$ pixels in the centre) are set to this pixel value threshold. 
After applying image scaling, we then normalised the images,
\begin{equation}
x=\frac{x-min(x)}{max(x)-min(x)},
\end{equation}
where $x$ is the pixel value. 
The effects of the various image scaling, for an example merger at $z=0.5$, can be seen in Fig. \ref{fig:scalings}.

These mock \textit{JWST}/NIRCAM images are used to train the merger classifier, as explained in the following section. It is generally expected that galaxy merging features are primarily driven by gravitational interactions, rather than the specific baryonic physics employed in a given simulation. A recent study \citep{2024MNRAS.530.4422K} focusing on tidal features fractions also qualitatively confirm this general expectation by showing a good level of agreement among four cosmological hydrodynamical simulations.



\section{Method}
\label{sec: method}

\subsection{Zoobot}
\label{sec: Zoobot}

\begin{table*}
\caption{Baseline network.}
    \centering
    \begin{tabular}{ |c|c|c|c|c| } 
    \hline
    Stage & Operator & Resolution & \#Channels & \#Layers \\
    \hline
    1 & Conv3x3 & 224 $\times$ 224 & 32 & 1 \\ 
    \hline
    2 & MBConv1, k3x3 & 112 $\times$ 112 & 16 & 1 \\ 
    \hline
    3 & MBConv6, k3x3 & 112 $\times$ 112 & 24 & 2 \\ 
    \hline
    4 & MBConv6, k5x5 & 56 $\times$ 56 & 40 & 2 \\ 
    \hline
    5 & MBConv6, k3x3 & 28 $\times$ 28 & 80 & 3 \\ 
    \hline
    6 & MBConv6, k5x5 & 14 $\times$ 14 & 112 & 3 \\ 
    \hline
    7 & MBConv6, k5x5 & 14 $\times$ 14 & 192 & 4 \\ 
    \hline
    8 & MBConv6, k3x3 & 7 $\times$ 7 & 320 & 1 \\ 
    \hline
    9 & Conv1x1 \& Pooling \& FC & 7 $\times$ 7 & 1280 & 1 \\ 
    \hline
    \end{tabular}
    \tablefoot{Architecture of the EfficientNet-B0 baseline network \citep{tan2020efficientnet}. The first layer is a convolutional layer of kernel size 3$\times$3. The majority of the model consists of mobile inverted bottleneck MBConv blocks. The last stage consists of a convolutional layer, a pooling layer and a fully connected (FC) layer.}
    \label{tab:effnet}
\end{table*}

We used Zoobot, a DL-based Python package \citep{Walmsley2023}, to build and optimise an image classifier to identify merging galaxies and identify their merging stages (i.e., pre- and post-mergers). The advantage of Zoobot is that it includes models that have already been pre-trained on the GZ responses. These models can be easily adapted to solve our task. 
The architecture of the pre-trained model used in this study is EfficientNet B0, which belongs to the family of models called EfficientNets \citep{tan2020efficientnet}. 
The baseline network consists of a convolutional layer of kernel size 3$\times$3, denoted by Conv3$\times$3. The majority of the model consists of mobile inverted bottleneck MBConv blocks, which uses an inverted residual block and a linear bottleneck. In the last stage there is a final convolutional layer, a pooling layer and a fully connected (FC) layer. The architecture of the network is summarised in Table \ref{tab:effnet}.

\begin{table*}
    \caption{The final parameters used for the training of each model.}
    \centering
    \begin{tabular}{|l|c|c|c|c|c|}
    \hline
       Classification set-up  & Batch size & Learning rate & Weight decay & Number of blocks & Scaling\\
       \hline
       One-stage & 128 & 1e-5 & 0.005 & 7 & aggressive arcsinh \\
       \hline
       Two-stage step 1 & 128 & 1e-5 & 0.05 & 4 & aggressive arcsinh\\
       \hline
       Two-stage step 2 & 128 & 1e-5 & 0.05 & 5 & aggressive arcsinh\\
       \hline
    \end{tabular}
    \tablefoot{The best hyperparameter values found for the models used in the one-stage and two-stage classification set-ups.}
    \label{tab:resulting parameters}
\end{table*}

The EfficientNet B0 model is trained on the GZ Evo dataset, as described in \citet{Walmsley2023}. This dataset comes from the GZ citizen science project where volunteers were asked questions based on morphological characteristics of galaxy images. There are various different versions of the project \citep{10.1093/mnras/stab2093}.
Questions from many versions are used to create the GZ Evo dataset consisting of 96.5M volunteer responses for 552k galaxy images. 
With this dataset, a pre-trained model is created, which is shown to have learned an internal representation capable of recognising general galaxy morphological types \citep{2022MNRAS.509.3966W}. Because of the generality of the representations, the resulting network is also suitable for solving new galaxy classification tasks via a method called transfer learning. 
As head this model uses a linear classifier, which applies a linear transformation to the input data given by, 
\begin{equation}
y=xA^T+b,
\end{equation}
where $x$ is the input data, $A$ is the weight matrix with the shape of the input and output features, $b$ is the bias and $y$ the output data. 
The output dimension is the number of classes (labels) that we consider in the classification task. 

To optimise the training on our own data, we first must obtain the best configuration of the tunable hyperparameters. In this work, we used the grid search approach which can be quite time-consuming. So carefully selecting which parameters and which settings to use is important. Therefore, we first did multiple episodes of random grid searches, where we provided multiple possible values from which random configurations were chosen. From the behaviour of each run, we learned more about its influences on the performance. Subsequently, for the final grid search, we were able to provide the most suitable parameters. 
The final grid search was performed over batch size, learning rate, weight decay, and the number of blocks that the optimiser will consider to update. 
We present the details of the grid search in Appendix \ref{appendix:tuning}. In Table \ref{tab:resulting parameters}, we summarise our findings for the best hyperparameter values to use for training our networks. 

\subsection{Performance metrics}
\label{sec: performance metrices}

In this study we make use of a multi-class classification model which predicts labels on input data. A common way to evaluate the performance of a classifier is to use the confusion matrix. Each row in a confusion matrix corresponds to the true class and each column corresponds to the predicted class. One can then use the confusion matrix to show the number of times instances of a given true class are classified as a given predicted class. For a perfect classifier, only the diagonal elements of the confusion matrix would have non-zero values. 
In order to properly compare the different classes in a confusion matrix, the result needs to be shown over a balanced test data set  (i.e., the number of instances is the same across the different true classes).

It is also common to normalise the counts in a confusion matrix. When normalising the confusion matrix vertically, each number in the matrix is divided by the sum of the corresponding column. In this case, per predicted class one can read off what percentage belongs to the true class and what percentage belongs to the other classes. The same can be done horizontally and represents the percentage of each predicted class for a given true class. 
By normalising vertically we can derive the precision (also known as purity) of each class along the diagonal line, which is given by
\begin{equation}
\label{eq: precision}
\text{precision}=\frac{\text{TP}}{\text{TP}+\text{FP}},
\end{equation}
where TP (true positives) refers to the number of instances that are correctly identified to be the class that we are interested in and FP (false positives) refers to the number of instances that are incorrectly identified to be the class that we are interested in. 
If we divide each element by the sum of the corresponding row, we get recall (also known as completeness) of each class along the diagonal,
\begin{equation}
\label{eq: recall}
    \text{recall}=\frac{\text{TP}}{\text{TP}+\text{FN}},
\end{equation}
where FN (false negatives) refers to the number of instances that are incorrectly identified to belong to the classes that we are not interested in.

Another useful metric is accuracy which relates the performance for all the classes together. In the binary classification scenario, accuracy can be calculated as below, 
\begin{equation}
\text{accuracy}=\frac{\text{TP+TN}}{\text{TP}+\text{TN}+\text{FP}+\text{FN}},
\end{equation}
where TN (true negatives) is the number of instances that are correctly identified to belong to the other classes. This definition can be easily adapted to a multi-class classification scenario.

\begin{figure*}
    \centering
    \includegraphics[width=7.4cm]{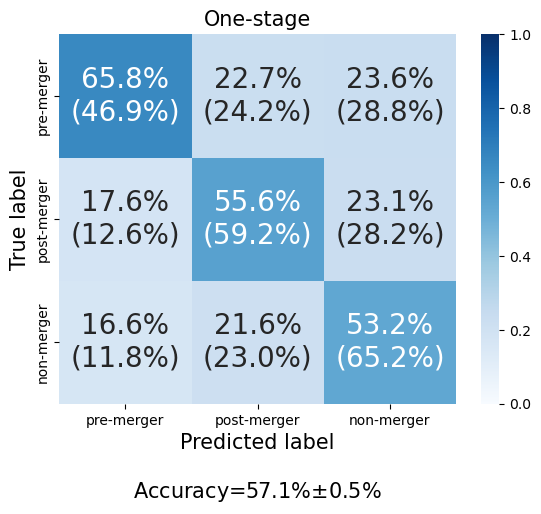}
    \includegraphics[width=7.4cm]{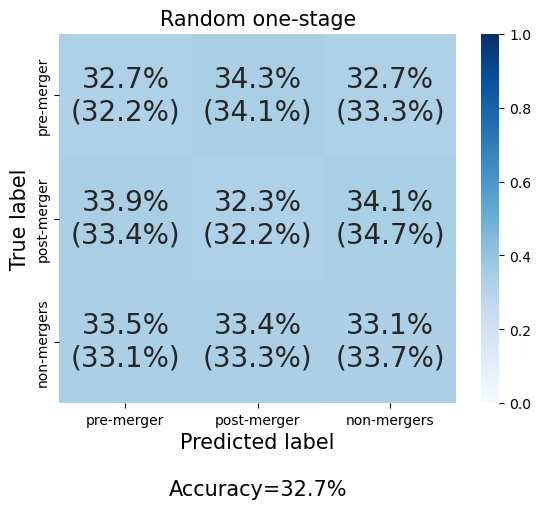}\\
    \vspace{0.3cm} 
    \includegraphics[width=7.4cm]{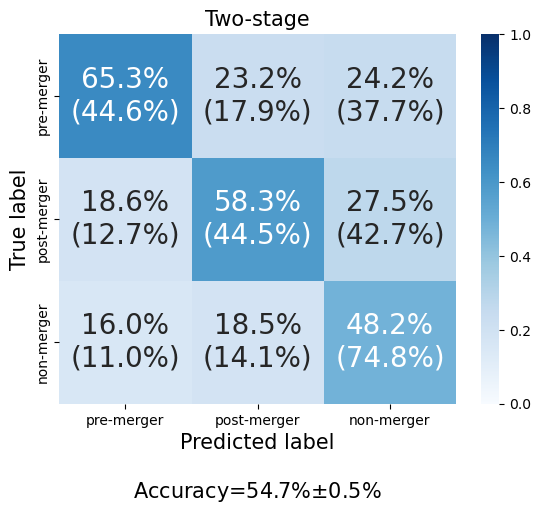}
    \includegraphics[width=7.4cm]{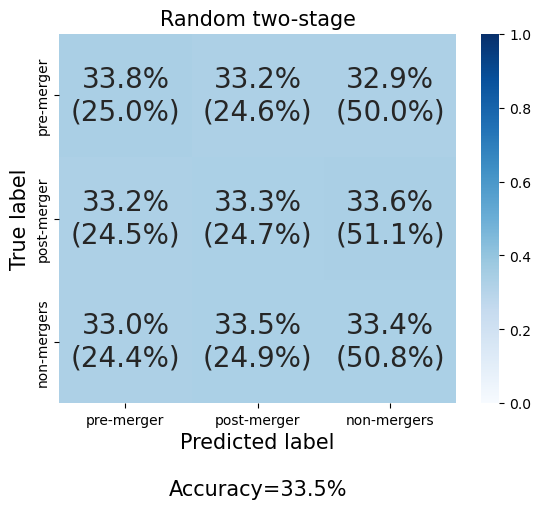}
    \caption{Top left: Confusion matrix for the one-stage classification, trained and predicted on the entire data set over all redshifts. The matrix is normalised vertically to give precision along the diagonal. Recall is shown in brackets below (i.e., the numbers in brackets are normalised horizontally). Top right: Example confusion matrix from a random 3-class classifier. Bottom left: Confusion matrix for the two-stage classification. Bottom right: Example confusion matrix from a random classifier in the two-stage  set-up. For clarity, only uncertainties on the accuracy metric are shown. Uncertainties on other metrics are broadly similar.}
    \label{fig:res all}
\end{figure*}

\begin{figure}
    \centering
    \includegraphics[height=7cm]{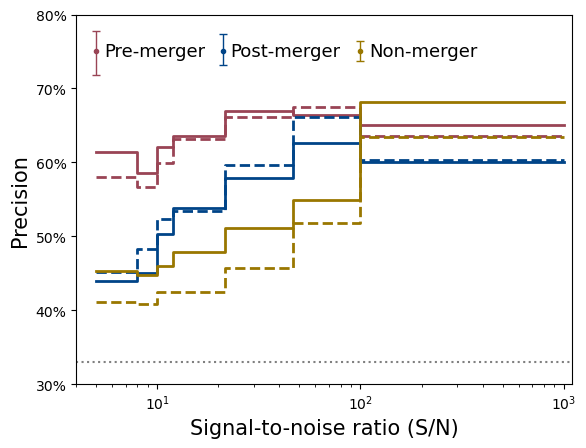}
    \caption{Precision as a function of  S/N for the three classes. The solid lines correspond to the one-stage classification and the dashed lines correspond to the two-stage classification. Precision levels increase with increasing S/N but these trends flatten at S/N $\gtrsim20$ for pre- and post-mergers. The mean errors on precision (derived from bootstrapping) are similar between the one-stage and two-stage set-ups and are indicated in the top left corner. The horizontal dotted line at around 33\% corresponds to the precision of a random classifier.}
    \label{fig:s_n_two}
\end{figure}

\begin{figure*}
    \centering
    \includegraphics[height=6cm, trim={0cm 0cm 2cm 0cm}, clip]{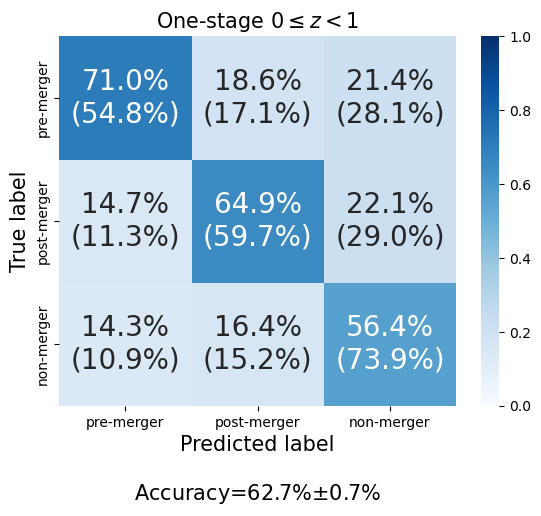}
    \includegraphics[height=6cm, trim={1.5cm 0cm 2cm 0cm}, clip]{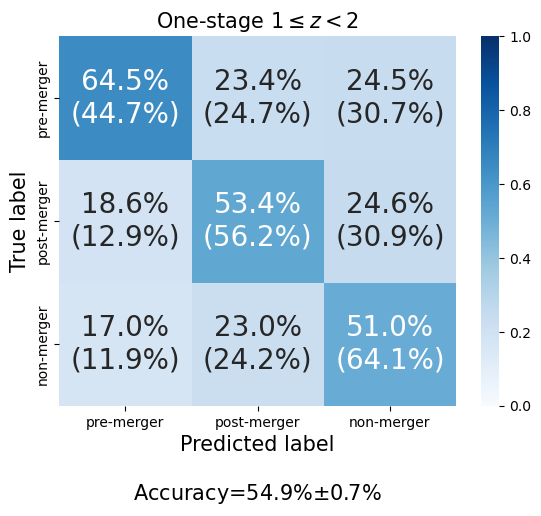}
    \includegraphics[height=6cm, trim={1.5cm 0cm 0cm 0cm}, clip]{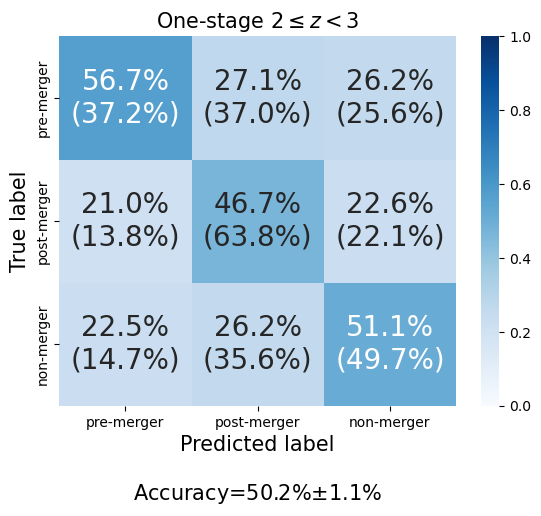}\\
    \vspace{0.3cm}
    \includegraphics[height=6cm, trim={0cm 0cm 2cm 0cm}, clip]{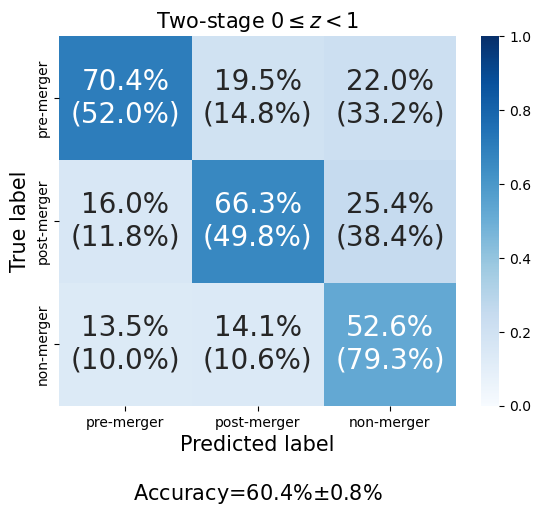}
    \includegraphics[height=6cm, trim={1.5cm 0cm 2cm 0cm}, clip]{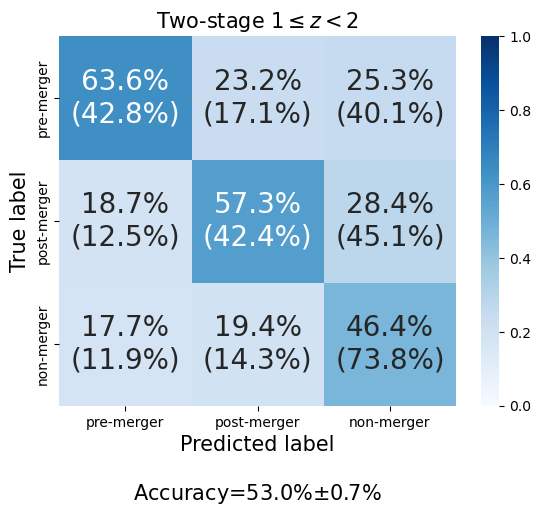}
    \includegraphics[height=6cm, trim={1.5cm 0cm 0cm 0cm}, clip]{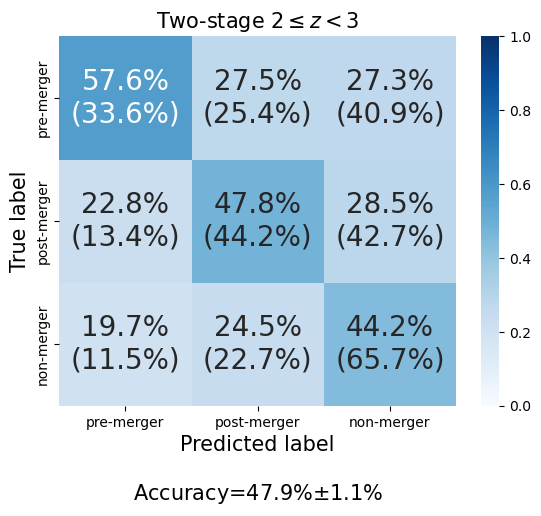}
    \caption{Confusion matrix of the one-stage (top row) and two-stage (bottom row)  predicted per redshift range (left: $0 \leq z < 1$; middle: $1 \leq z < 2$; right: $2 \leq z < 3$). All matrices are vertically normalised. Recall is shown below in brackets.}
    \label{fig:two_red_pred}
\end{figure*}

\section{Results}
\label{sect: results}

We train DL CNN models to identify mergers and merger stages, using mock {\it JWST} NIRcam F150W images of galaxies selected from the TNG simulations. Standard data augmentation techniques, such as cropping and zooming, random rotations, and random horizontal and vertical flipping, are included in Zoobot by default. We use two different set-ups. In the first set-up, we adopt a two-stage classification, in which the first step performs a binary classification of mergers and non-mergers. The predicted mergers then go through the second step which performs a binary classification of pre-mergers and post-mergers. 
The combined results from the two steps give the final three-class classifications of pre-merger, post-merger and non-merger. In the second set-up, the three classes (i.e., non-mergers, pre-mergers and post-mergers) are predicted simultaneously, which is referred to as the one-stage classification. 


For the first model of the two-stage classification set-up, we used a dataset of 350\,133 mergers and non-mergers of which $81\%$ is used for training, $9\%$ for validating and $10\%$ for testing. For the second model of the two-stage classification, 461\,571 pre- and post-mergers are used and the same training-validation-testing split is applied. For the one-stage classification set-up, we used 499\,824 galaxies in total, split in $81\%$ for training, $9\%$ for validating and $10\%$ for testing. A full breakdown of the number statistics can be found in Table \ref{tab:split}.

\subsection{Overall performance}

\begin{table*}
\caption{Number statistics  for the training of each model.}
    \centering
    \begin{tabular}{|l|c|c|c|c|c|c|}
    \hline
Classification set-up & Dataset & Mergers & Non-mergers & Pre-mergers & Post-mergers & Total \\
\hline
\multirow{3}{*}{Two-stage step 1} & Training & 147\,672 & 134\,763 & - & - & 282\,435 (81\%) \\
& Validation & 16\,911 & 14\,911 & - & - & 31\,902 (9\%) \\
& Test & 18\,942 & 16\,854 & - & - & 35\,796 (10\%) \\
\hline
\multirow{3}{*}{Two-stage step 2} & Training & - & - & 186\,822 & 184\,728 & 371\,550 (80\%) \\
& Validation & - & - & 21\,570 & 20\,883 & 42\,453 (9\%) \\
& Test & - & - & 24\,096 & 23\,472 & 47\,568 (10\%) \\
\hline
\multirow{3}{*}{One-stage} & Training & - & 134\,763 & 134\,763 & 134\,763 & 404\,289 (81\%) \\
& Validation & - & 14\,991 & 14\,991 & 14\,991 & 44\,973 (9\%) \\
& Test & - & 16\,854 & 16\,854 & 16\,854 & 50\,562 (10\%)\\

\hline
\end{tabular}
    \tablefoot{Percentages are shown in brackets.}
    \label{tab:split}
\end{table*}

\begin{table*}
    \caption{Number of classes in each training dataset  per signal-to-noise ratio (S/N) bin.}
    \centering
    \begin{tabular}{|l|c|c|c|c|}
    \hline
      S/N  & Mergers & Non-mergers & Pre-mergers & Post-mergers \\
      \hline
      5-8  & 17\,343 & 18\,435 & 19\,236 & 26\,004 \\
      \hline
      8-10 & 9\,216 & 10\,281 & 10\,638 & 13\,178  \\
      \hline
      10-12 & 7\,329 & 9\,018 & 9\,024 & 9\,567 \\
      \hline
      12-21.5 & 23\,796 & 27\,651 & 29\,268 & 31\,080\\
      \hline
      21.5-46.4 & 28\,341 & 29\,244 & 37\,260 & 32\,958 \\
      \hline
      46.4-100 & 26\,565 & 22\,446 & 36\,636 & 28\,170 \\
      \hline
      100-1000 & 40\,014 & 12\,639 & 44\,694 & 43\,539\\
      \hline
    \end{tabular}
    \tablefoot{We chose roughly equal number of galaxies in each of the classes. However, the last bin is highly unbalanced, with significantly fewer non-mergers.}
    \label{tab:N galaxies per S/N ratio}
\end{table*}


\paragraph{One-stage classification}
First, we show the results of the one-stage classification on the three classes, pre-merger, post-merger and non-mergers. From the grid search, we obtained the best values for the tuned hyper-parameters, as shown in Table \ref{tab:resulting parameters}. Using these parameters we trained our models on the IllustrisTNG dataset, which consists of 134\,763 galaxies for each of the three classes. The resulting confusion matrix from applying the trained model to the test dataset (balanced over the three classes) is shown in the top left panel in Fig. \ref{fig:res all}. 
Galaxies are separated into different classes based on the highest prediction score from the model. 
The overall accuracy of our one-stage classification set-up is 57.1\% ($\pm0.5\%$).
The precision for the pre-merger class is 65.8\%, much higher than the other two classes which show precisions around 55\%. We can postulate that this is probably due to the generally more recognisable and distinct features in pre-mergers, e.g., distortions in the primary galaxy, the presence of a merging companion, and interacting morphologies between the merging pair. 
On the other hand, recall is the lowest for pre-mergers ($<50\%$) and much higher for the other classes ($\sim60$\% or higher). This is not surprising given the well-known trade-off between precision and recall. Mis-classifications of pre-mergers are roughly evenly distributed over the other two classes, at a level of around 25\%. For post-mergers and non-mergers, confusion with the pre-merger class (at just over 10\%) is significantly lower than with each other by around a factor of two. This is likely due to the fact that post-mergers, like non-mergers, also do not have a merging companion galaxy. Indeed, several previous studies \citep[e.g., ][]{2019A&A...626A..49P, 2020A&C....3200390C, 2024arXiv240718238L} have investigated which specific image features can lead to a merger classification, by using techniques such as occlusion analysis and Gradient-weighted Class Activation Maps \citep{2016arXiv161002391S}. Generally speaking, models rely on the presence of a merging companion and/or complex asymmetric (usually faint) features around the periphery of the central galaxy to distinguish mergers from non-mergers. To help interpret our results further, we also plot a confusion matrix corresponding to a random 3-class classifier in the same set-up in the top right panel in Fig. \ref{fig:res all}. When classifications are done randomly, each class has an equal probability of being predicted and hence, this results in precision and recall scores of approximately 33\% for all classes and an overall accuracy of around 33\%. In Appendix \ref{appendix:binary}, we also directly compare the performance of our classifiers in the binary classification case (mergers vs. non-mergers) with other leading merger detection methods as presented in \citet{2024A&A...687A..24M}, demonstrating overall similar performance levels.



\paragraph{Two-stage classification}
A similar approach was used for the two-stage classification set-up, but the grid search was done for the two steps separately. The best hyper-parameter configurations obtained are shown in Table \ref{tab:resulting parameters}. The configuration for the model used in the second step (i.e., the pre-merger/post-merger classification) is the same as the one used in the first step (i.e., the merger/non-merger classification), except that we fine-tuned one additional deeper block. Furthermore, we applied class weights to both models as the data are slightly unbalanced. The model used in the first step was trained on 147\,672 mergers and 134\,763 non-mergers. In the second step, initially, the two classes were fairly unbalanced, with  pre-mergers being more numerous than post-mergers. To balance the datasets and to increase the training data, we performed data augmentation for both classes, resulting in 186\,822 pre-mergers and 184\,728 post-mergers. The augmentations were done by producing new mock images from the simulated galaxy images with different backgrounds. 
Our test dataset was first subjected to a merger/non-merger classification. Galaxies predicted as mergers were further classified into pre- and post-mergers. The results are shown in the bottom left panel in Fig. \ref{fig:res all}. The overall accuracy is 54.7\%   ($\pm0.5\%$), slightly lower than that of the one-stage set-up.  The precisions of the three classes are either similar or lower than the ones in the one-stage classification, particularly for the non-merger class (5\% lower). Again, we see more confusion between the non-mergers and post-mergers, compared to confusions between these two classes and the pre-merger class. 
In the bottom right panel in Fig. \ref{fig:res all}, we also show the results of a random classifier in the two-stage set-up. In this case, a random classifier would still give an overall accuracy of around 33\% and precisions $\sim33\%$ for all classes, but the recall would be around 50\% for non-mergers and 25\% for pre- and post-mergers (as the test dataset is balanced over pre-mergers, post-mergers and non-mergers).

\begin{table*}
\center
\caption{Precision and recall score of the two steps of the two-stage classification. }
  \begin{tabular}{|l|l|l|l|l|l|l|}
    \hline
      \multirow{2}{*}{Two-stage classification} &
      \multicolumn{3}{c|}{Precision} &
      \multicolumn{3}{c|}{Recall} \\\cline{2-7}
     & $0\leq z<1$ & $1\leq z<2$ & $2\leq z<3$ & $0\leq z<1$ & $1\leq z<2$ & $2\leq z<3$ \\
    \hline
    Step 1 non-merger & 69\% & 63\% & 62\% & 80\% & 73\% & 66\% \\
    \hline
    Step 1 merger & 76\% & 69\% & 64\% & 64\% & 57\% & 60\% \\
    \hline
    Step 2 pre-merger & 80\% & 76\% & 72\% & 78\% & 71\% & 57\% \\
    \hline
    Step 2 post-merger & 79\% & 73\% & 64\% & 81\% & 78\% & 77\% \\
    \hline
  \end{tabular}
  \tablefoot{ 
  The first step is the merger/non-merger classification. The second step is the pre-merger/post-merger classification for which only the true positives from the first step are considered. 
  }
  \label{tab: redshift perf two}
\end{table*}

\begin{figure*}
    \centering
 \includegraphics[width=\linewidth]{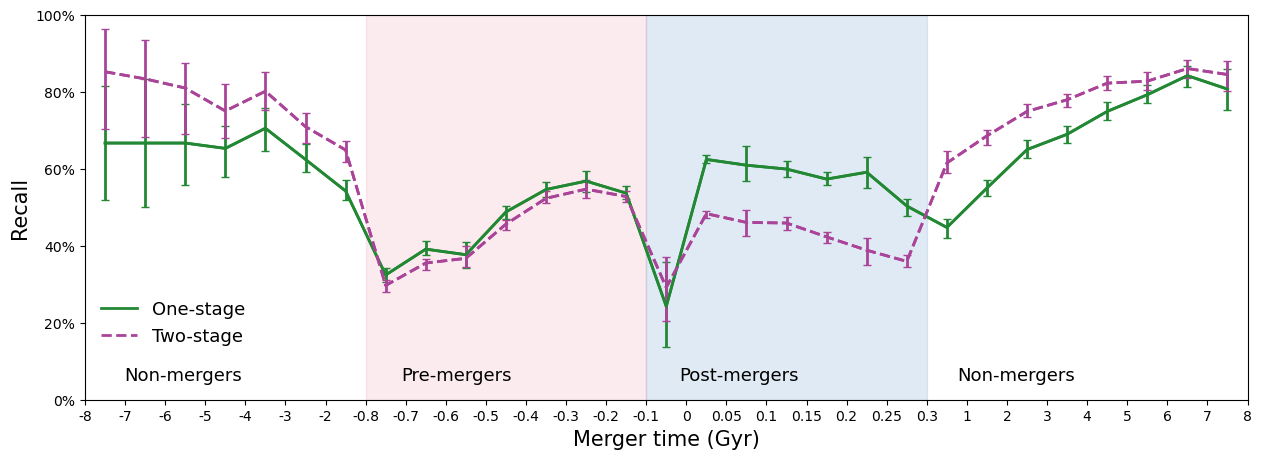}   
    \caption{Recall of each class as a function of merger time before or after the coalescence (one-stage: green solid line; two-stage: purple dashed line). Note that the x-axis is stretched in the merger regime.  The blue-shaded area corresponds to the true post-merger regime, the pink-shaded area corresponds to the true pre-merger region, and the white area is the true non-merger regime. Error bars are derived from bootstrap error.}
    \label{fig:comp rec 1 vs 2}
\end{figure*}

\paragraph{Dependence on S/N} We now analyse how S/N affects the precision of the two classification set-ups. 
We use \verb|photutils| package \citep{larry_bradley_2024_12585239} to estimate the aperture photometry (within a circular aperture of 6 pixels in diameter, i.e., around 0.2\arcsec) and the background noise level within the same size area (as the sigma-clipped standard deviation), from which we calculate the S/N.
All galaxies in the dataset were binned according to their S/N over the range from 5 to 1000. 
We randomly chose in each S/N bin a roughly equal number of galaxies from each class, listed in Table \ref{tab:N galaxies per S/N ratio}. The bins up to a S/N of 100 contain approximately balanced classes between mergers and non-mergers. However, the last bin is highly unbalanced, with significantly fewer non-mergers.
Next, we calculate the precision of each class in each S/N bin. The results are plotted in Fig. \ref{fig:s_n_two}, where each horizontal line span across the width of the S/N bin. 
In general, the precision levels of all three classes increase as the S/N increases for both the one-stage and two-stage set-ups, as expected. At S/N $\gtrsim20$, the trends for pre- and post-mergers become much flatter, indicating much less dependence on S/N. For the non-merger class, precision continues to rise sharply with increasing S/N.
This could be partly due to the imbalance of the three classes in the highest S/N bin which can lead to a bias in model predictions towards predicting merging galaxies, resulting in a lower precision for both pre- and post-mergers. 
At any S/N bin except for the highest bin, pre-mergers have the highest precision, non-mergers have the lowest precision and post-mergers are located in-between, for both set-ups. 


\paragraph{Dependence on redshift} To examine the dependence of the model performance on redshift, we show in Fig. \ref{fig:two_red_pred} the confusion matrices of the model trained on all redshifts, but predicted separately for each redshift bin. 
As expected, in both set-ups predicting on lower-$z$ galaxies generally gives higher precision, recall and accuracy for all  classes, as typically lower-$z$ galaxies are larger and have higher S/N. 
Again, the one-stage set-up generally has better precision, particularly for non-mergers. 
Since the two-stage classification set-up is dependent on two steps, we  also present the performance of the two separate models per redshift range in Table \ref{tab: redshift perf two}. 
In the first step, the recall of the non-merger class decreases much quicker than the merger-class. At $0 \leq z < 1$, the recall is $80\%$ and $64\%$ for the non-mergers and mergers, respectively. At $2 \leq z < 3$, the recall has reduced by $14\%$ for the non-mergers and only $4\%$ for the mergers. On the other hand, the precision of the mergers is much more affected by redshift than that of the non-mergers.
This ultimately results in more FP (i.e., non-mergers classified as mergers) at higher-$z$ and therefore contaminating the second step of the classification. 
To analyse the results from the second step, we only consider the TPs (i.e., correctly identified mergers) from the first step. 
A significant decline in recall is only found in the pre-mergers. However, precision declines more sharply for the post-mergers. As discussed previously, this could be partially due to the number distribution of the different classes in the high-$z$ bin (lowest S/N bin) which is mainly dominated by post-mergers. In a separate experiment, we also trained the classifiers in each redshift bin separately. However, as the numbers of galaxies used for training decreased by a factor of 10 in the highest redshift bin, the results did not show much improvement.

To summarise, the one-stage classification set-up moderately outperforms the two-stage set-up as it offers better accuracy and precision. The recall of the one-stage set-up for the non-merger class is worse than the two-stage set-up. However, precision is usually more important than recall for merger studies. In addition, mergers are relatively rare and so a lower recall in the non-mergers class is not an issue as far as sample size is concerned. One possible reason why the one-stage set-up outperforms the two-stage set-up is because it can be more difficult for DL models to identify a class that consists of a mix of morphologies types. A set-up that puts more similar galaxies together in a finer classification system can be helpful for the DL models to learn the (sometimes very) subtle differences between the various classes better.

\begin{figure*}
    \centering
    \includegraphics[width=\linewidth]{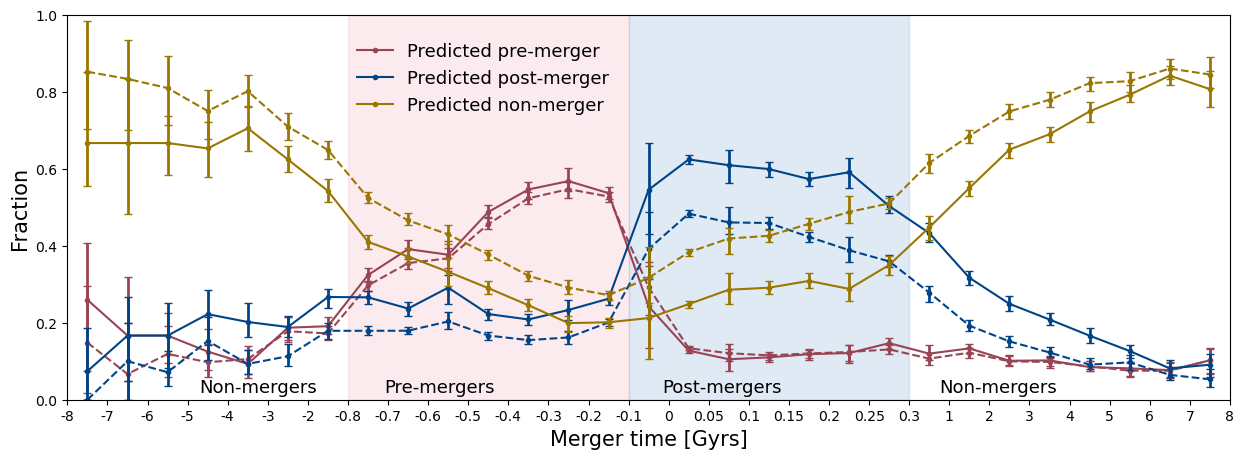}
    \caption{Fraction of the three predicted classes as a function of merger time, before or after coalescence, in the regimes of true mergers (blue-shaded region: true post-mergers; pink-shaded region: true pre-mergers) and true non-mergers. The x-axis is stretched in the merger regime.  The solid lines indicate the results of the one-stage classification set-up. The dashed lines indicate the results of the two-stage classification set-up. Error bars are derived from bootstrap error. For a perfect classifier, the fraction of predicted pre-mergers/post-mergers/non-mergers would be 100\% in the true pre-merger/post-merger/non-merger region and 0\% everywhere else.} 
    \label{fig:compare frac}
\end{figure*}

\begin{figure*}
    \centering
    \includegraphics[width=\linewidth]{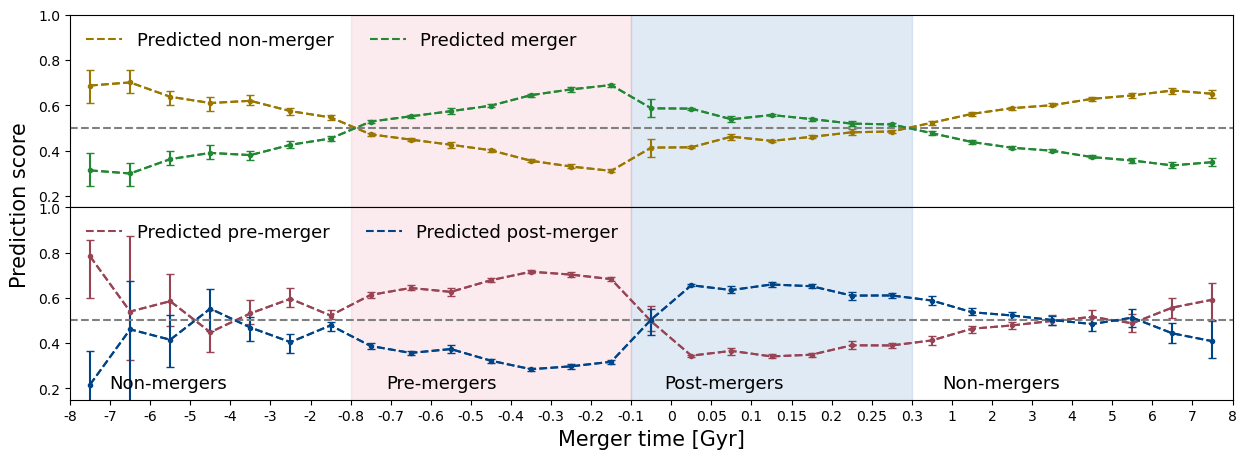}
    \includegraphics[width=\linewidth]{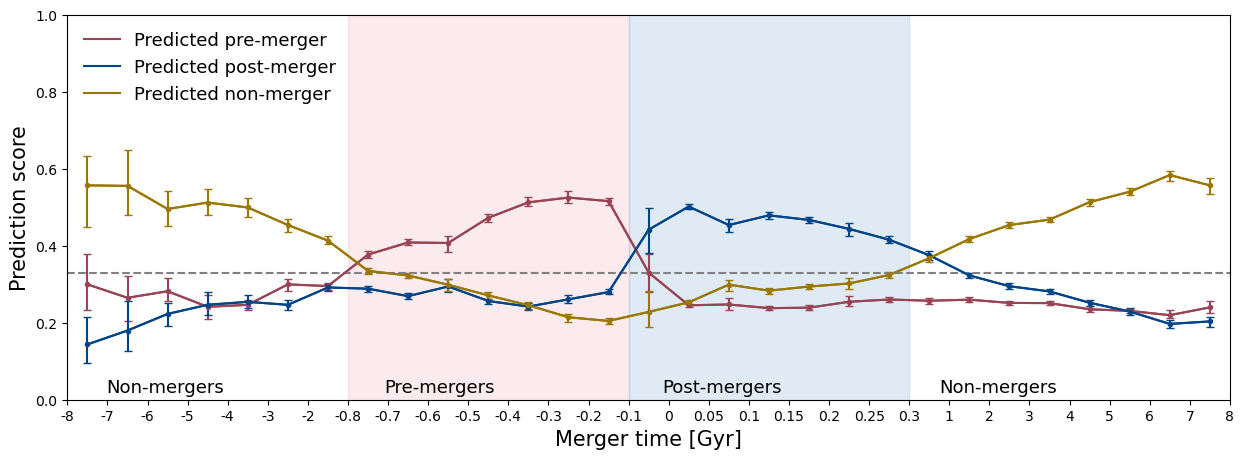}\\
    \caption{Prediction score  as a function of merger time (top panel: the two-stage classification set-up; bottom panel: the one-stage set-up). The x-axis is stretched in the merger regime.  In the two-stage set-up, the first stage gives each galaxy a merger (green) and non-merger (yellow) prediction score. For the predicted mergers, the second stage gives a pre-merger (purple) and post-merger (blue) prediction score. The error bars correspond to bootstrap errors. For the one-stage/two-stage set-up a grey dotted line at $0.33$/$0.5$ is added to indicate the prediction score of a random classifier.}
    \label{fig: comp prediction}
\end{figure*}

\subsection{Dependence on merger time}
\label{sec: Dependencies}

During the Gyr-long galaxy merging processes, merging features clearly evolve as the galaxies interact with each other. Some merging stages associated with clear and distinct signs such as tidal tails and bridges can be easier to identify than other stages. 
We want to understand how merger time $t_{\rm merger}$ before or after the coalescence event affects predictions scores of each class, with a focus at the boundaries between the different classes.

Fig. \ref{fig:comp rec 1 vs 2} shows the recall of the three classes against merger time $t_{\rm merger}$ for both set-ups. The pink (blue) shaded region corresponds to the regime of the true pre-mergers (post-mergers) and the white regions correspond to the regime of the true non-mergers.
In general, the recall levels for both pre- and post-mergers increase when those mergers are closer to the coalescence event, indicating that it is easier to detect them possibly due to more conspicuous features. 
At the boundary between pre-mergers and post-mergers, recall is significantly decreased which is expected as there is increased confusion between the two classes at the boundary.
For non-mergers, recall increases for galaxies with more distant merging events in their past or future evolutionary histories, as those galaxies are likely to have more regular morphologies. The one-stage classification set-up results in higher recall for mergers, particularly for post-mergers. 
In the two-stage set-up, the two merger classes are predicted via two networks and have therefore a higher chance of being polluted by other classes. 
For non-mergers, the two-stage set-up has higher recall than the one-stage set-up. This is to be expected as the precision of the non-mergers in the two-stage set-up is worse than in the one-stage set-up. Another point to keep in mind is that the recall of a random classifier in the two-stage set-up is around 50\%. Therefore, the overall recall of the non-merger class in the two-stage set-up, at around 75\% (as shown in Fig. \ref{fig:res all}),  is 25\% points better than random classifications. In comparison, the overall recall of the non-merger class in the one-stage set-up, at around 65\%, is $\sim30\%$ better compared to that from a random classifier in the one-stage set-up.

In Fig. \ref{fig:compare frac}, we plot the fraction of each predicted class versus $t_{\rm merger}$.
In the post-merger regime, a perfect classifier would return 100\% for the predicted post-mergers and 0\% for the predicted pre-mergers and non-mergers. For both the one-stage and two-stage set-ups, the fraction of the predicted post-mergers decreases with increasing merger time after coalescence. This indicates that the classifiers are more confused, probably due to the vanishing merging features.
For the two-stage set-up, galaxies are even more likely to be classified as non-mergers at $t_{\rm merger}>0.15$ Gyr. 
In comparison, in the one-stage set-up, the fraction of the predicted post-mergers is much higher and the fraction of the predicted non-mergers is much lower, implying this set-up is better at differentiating post-mergers and non-mergers.
In the pre-merger regime, both set-ups show similar fractions of predicted pre-mergers, which generally decrease with increasingly negative merger times (coalescence is further away in the future). 
Both set-ups perform well in the region from around $t_{\rm merger} = -0.2$ up to $-0.5$ Gyr. Beyond $t_{\rm merger} = -0.5$ Gyr, the fraction of predicted non-mergers is higher than that of the predicted pre-mergers in the two-stage set-up. In comparison, the fraction of the predicted non-mergers does not overtake that of the predicted pre-mergers until around $t_{\rm merger} = -0.7$ Gyr in the one-stage set-up. 
In the non-merger regime, the two-stage set-up returns a higher fraction of predicted non-mergers than the one-stage set-up. This is related to the fact that the one-stage set-up has higher precision for the non-merger class, which generally means recall will be lower.

We now investigate how confident the classifier is as a function of $t_{\rm merger}$, which is reflected in the prediction scores that the model gives to each class. For each galaxy, the highest score will determine the predicted label. If the model is very confident, the prediction score of a particular class will be much higher than for the other classes. 
In Fig. \ref{fig: comp prediction}, we plot the average pre-, post-, and non-merger prediction scores as a function of $t_{\rm merger}$. 
For the two-stage set-up, the first model gives a merger  score and non-merger score. 
The second model gives a pre-merger score and post-merger score, which are only given to galaxies with a merger score $>0.5$ from the previous stage. As such, non-mergers predicted as mergers in the first stage will be classified as pre- or post-merger in the second stage. Hence, one of these scores could be very high, even though the original merger score was perhaps fairly low. 
For the two-stage set-up, the highest average score for a merger is in the true pre-merger regime, particularly in the range $t_{\rm merger}=-0.5$ to $-0.1$ Gyr. 
The merger scores are generally significantly lower in the true post-merger regime than in the true pre-merger regime. This implies that for post-mergers, the first stage is the main problem for mis-classifications. 
This confusion is much less in the one-stage classification set-up. 
Comparing the true non-merger regimes in both set-ups, the non-merger prediction score decreases towards the boundaries with true pre- and post-mergers. 

\subsection{Retraining the network}

\begin{figure*}
    \centering
    \includegraphics[width=0.48\linewidth]{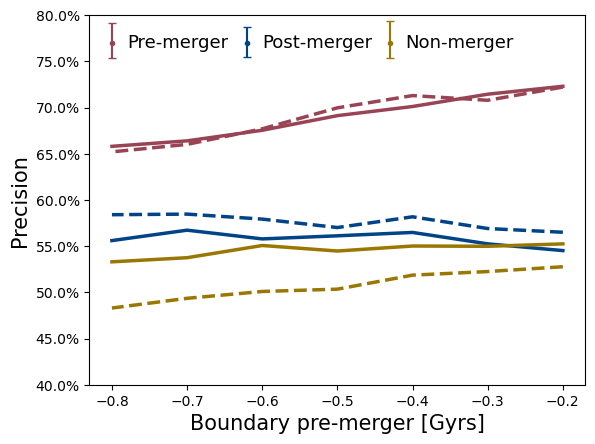}
\includegraphics[width=0.48\linewidth]{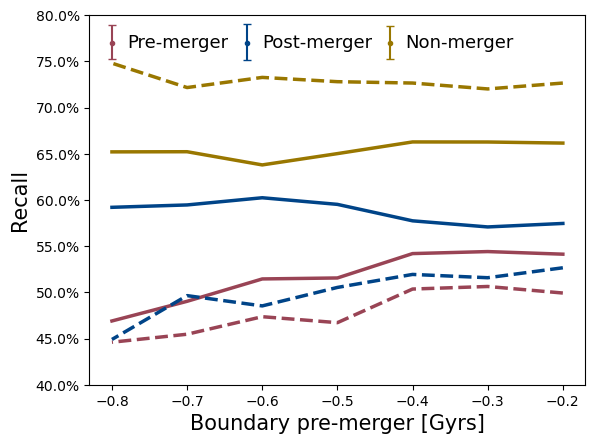}
    \caption{Precision (left) and recall (right) scores of the three classes as a function of the new pre-merger definition. The solid lines represent the one-stage classification and the dashed lines represent the two-stage classification. The error bars derived from bootstrapping are similar between the one-stage and two-stage set-ups. The mean errors for the three classes are indicated in the top. }
    \label{fig:performance tb m1}
\end{figure*}

In Sect. \ref{sec: Dependencies}, we showed that the performance metrics  of the classifiers and the confidence levels of their predictions depend sensitively on the merger time $t_{\rm merger}$ since or before the coalescence. In general, the classifiers give better precision, recall and higher confidence for mergers which are caught closer to the actual merging event. In our default definition of pre-mergers, we used a relatively long time range between $t_{\rm merger}=$ -0.8 and -0.1 Gyrs. Therefore, for the pre-merger class we would like to check if it is better to pick a different boundary, by changing the definition of the pre-merger and non-merger class and then retrained the two classification set-ups for each new definition. 

We considered new boundaries at $t_{\rm merger}= -0.7, -0.6, -0.5, -0.4, -0.3, -0.2, $ and $-0.1$ Gyrs. Pre-mergers outside the new definition window are now labelled as non-mergers. 
For example, a pre-merger which is located along the merging sequence at $t_{\rm merger}=-0.65$ Gyrs is initially defined as a pre-merger. When we changed the definition of pre-mergers using a new boundary at $t_{\rm merger}=-0.6$ Gyrs, this galaxy would then be labelled as a non-merger. 
In terms of the overall accuracy for the three classes, we find that the highest overall accuracy for the two-stage (one-stage) set-up, $58.4\%$ ($59.4\%$), is achieved when $t_{\rm merger}=0.2$ (0.4) Gyrs is adopted. This finding supports our previous conclusion that the one-stage set-up generally performs better than the two-stage set-up, with better accuracy and larger range in merger time.  

In Fig \ref{fig:performance tb m1}, we plot the precision and recall as a function of the boundary set between pre-mergers and non-mergers, for both classification set-ups. 
For both metrics, pre-mergers are clearly the most affected when changing the boundary, which is expected as the other two classes (post-mergers and non-mergers) are not much affected.
The precision levels of the pre-mergers continues to rise as we focus on pre-mergers which are located closer to the actual merging event, for both classification set-ups, from around 65\% at $t_{\rm merger} = -0.8$ Gyr (the original definition) to around 72\% at $t_{\rm merger} = -0.2$ Gyr.
Similarly, the recall levels of the pre-mergers also generally improve from $t_{\rm merger} = -0.8$ to $t_{\rm merger} = -0.2$ Gyr but these trends flatten more or less at around $t_{\rm merger} = -0.4$ Gyr. 
When taking into account all the classes the best definition of pre-merger would be galaxies that will under go a merger event in the following 0.4 Gyr. The results from the model trained with this definition showed improvement for the majority of classes in terms of both precision and recall.

\section{Conclusion}
\label{sect: conclusion}

In this paper, we use advanced DL models as implemented in the Zoobot Python package to detect non-mergers and mergers including their merger stages in mock {\it JWST}/NIRCam F150W images of simulated galaxies over the redshift range $0.5\leq z\leq 3.0$ selected from the IllustrisTNG cosmological hydrodynamical simulations. We employ two classification set-ups, i.e. one-stage vs two-stage classification, to test if it is more advantageous to detect mergers and merger stages in a hierarchical fashion or simultaneously. We also investigate how performance of our classifiers change as a function of signal-to-noise, redshift, and merger time relative to the merging event. Our main findings are the following:
\begin{itemize}
        \item The one-stage classification set-up which identifies mergers and their merger stages simultaneously generally performs better than the hierarchical two-stage set-up, yielding higher overall accuracy and generally better precision (particularly for the non-merger class). In terms of recall, the one-stage set-up performs better for pre- and post-mergers, but worse for non-mergers. However, this is partly due to the trade-off between precision and recall. For mergers studies, we generally prioritise precision over recall. Therefore, we recommend the one-stage set-up for identification of mergers and their merger stages.
        \item Pre-mergers are classified with the highest precision. There is more confusion between the post-mergers and non-mergers than between these two classes and the pre-merger class. This is probably not surprising as pre-mergers can have more recognisable features, e.g., the presence of a merging companion galaxy in addition to the distorted morphologies in the primary galaxy.
        \item Precision of the DL classifiers generally improve with increasing S/N. The trends are particularly steep at the faint end and then flatten towards higher S/N. Therefore, it is important for studies involving merger detections to investigate first the dependence on S/N and then choose an optimal threshold to balance between performance and sample size.  
        \item The performance of the classifiers depends on the exact merger stage in terms of time since or before the merging event (i.e., coalescence). Precision continues to increase for pre-mergers closer to coalescence, but recall plateaus at around $t_{\rm merger}=-0.4$ Gyr. To keep a good balance between precision and recall, also taking into account consideration of sample size, we therefore recommend using a cut at -0.4 Gyr to select pre-mergers.
\end{itemize}

In our future work, we will explore the connection between key galaxy evolutionary phases (such as the triggering of starburst activities and accretion onto SMBHs) and merger stages in real \textit{JWST} data. For example, we will address if the starburst phase is more likely to be activated in the pre-merger stage and the SMBH accretion more likely to be triggered in the post-merger stage, using the one-stage classification set-up and the recommended cuts on S/N and merger time from this paper. In this paper, we investigated merger stages at the most basic level, i.e., separating mergers into pre- and post-mergers. Another future direction would be to refine the time resolution in pinpointing mergers along their merging sequences.


\begin{acknowledgements}

This publication is part of the project `Clash of the Titans: deciphering the enigmatic role of cosmic collisions' (with project number VI.Vidi.193.113 of the research programme Vidi which is (partly) financed by the Dutch Research Council (NWO). 
We thank the Center for Information Technology of the University of Groningen for their support and for providing access to the Hábrók high-performance computing cluster. We thank SURF (www.surf.nl) for the support in using the National Supercomputer Snellius.
W.J.P. has been supported by the Polish National Science Center project UMO-2023/51/D/ST9/00147.
The Dunlap Institute is funded through an endowment established by the David Dunlap family and the University of Toronto.
\end{acknowledgements}

\bibliography{bibliography_rosa}

\begin{thebibliography}{90}
\expandafter\ifx\csname natexlab\endcsname\relax\def\natexlab#1{#1}\fi

\bibitem[{{Ackermann} {et~al.}(2018){Ackermann}, {Schawinski}, {Zhang}, {Weigel}, \& {Turp}}]{2018MNRAS.479..415A}
{Ackermann}, S., {Schawinski}, K., {Zhang}, C., {Weigel}, A.~K., \& {Turp}, M.~D. 2018, \mnras, 479, 415

\bibitem[{{Aihara} {et~al.}(2018){Aihara}, {Arimoto}, {Armstrong}, {Arnouts}, {Bahcall}, {Bickerton}, {Bosch}, {Bundy}, {Capak}, {Chan}, {Chiba}, {Coupon}, {Egami}, {Enoki}, {Finet}, {Fujimori}, {Fujimoto}, {Furusawa}, {Furusawa}, {Goto}, {Goulding}, {Greco}, {Greene}, {Gunn}, {Hamana}, {Harikane}, {Hashimoto}, {Hattori}, {Hayashi}, {Hayashi}, {He{\l}miniak}, {Higuchi}, {Hikage}, {Ho}, {Hsieh}, {Huang}, {Huang}, {Ikeda}, {Imanishi}, {Inoue}, {Iwasawa}, {Iwata}, {Jaelani}, {Jian}, {Kamata}, {Karoji}, {Kashikawa}, {Katayama}, {Kawanomoto}, {Kayo}, {Koda}, {Koike}, {Kojima}, {Komiyama}, {Konno}, {Koshida}, {Koyama}, {Kusakabe}, {Leauthaud}, {Lee}, {Lin}, {Lin}, {Lupton}, {Mandelbaum}, {Matsuoka}, {Medezinski}, {Mineo}, {Miyama}, {Miyatake}, {Miyazaki}, {Momose}, {More}, {More}, {Moritani}, {Moriya}, {Morokuma}, {Mukae}, {Murata}, {Murayama}, {Nagao}, {Nakata}, {Niida}, {Niikura}, {Nishizawa}, {Obuchi}, {Oguri}, {Oishi}, {Okabe}, {Okamoto}, {Okura}, {Ono}, {Onodera}, {Onoue}, {Osato}, {Ouchi}, {Price}, {Pyo},
  {Sako}, {Sawicki}, {Shibuya}, {Shimasaku}, {Shimono}, {Shirasaki}, {Silverman}, {Simet}, {Speagle}, {Spergel}, {Strauss}, {Sugahara}, {Sugiyama}, {Suto}, {Suyu}, {Suzuki}, {Tait}, {Takada}, {Takata}, {Tamura}, {Tanaka}, {Tanaka}, {Tanaka}, {Tanaka}, {Terai}, {Terashima}, {Toba}, {Tominaga}, {Toshikawa}, {Turner}, {Uchida}, {Uchiyama}, {Umetsu}, {Uraguchi}, {Urata}, {Usuda}, {Utsumi}, {Wang}, {Wang}, {Wong}, {Yabe}, {Yamada}, {Yamanoi}, {Yasuda}, {Yeh}, {Yonehara}, \& {Yuma}}]{2018PASJ...70S...4A}
{Aihara}, H., {Arimoto}, N., {Armstrong}, R., {et~al.} 2018, \pasj, 70, S4

\bibitem[{{Bickley} {et~al.}(2021){Bickley}, {Bottrell}, {Hani}, {Ellison}, {Teimoorinia}, {Yi}, {Wilkinson}, {Gwyn}, \& {Hudson}}]{2021MNRAS.504..372B}
{Bickley}, R.~W., {Bottrell}, C., {Hani}, M.~H., {et~al.} 2021, \mnras, 504, 372

\bibitem[{Biewald(2020)}]{wandb}
Biewald, L. 2020, Experiment Tracking with Weights and Biases, software available from wandb.com

\bibitem[{{Blecha} {et~al.}(2018){Blecha}, {Snyder}, {Satyapal}, \& {Ellison}}]{2018MNRAS.478.3056B}
{Blecha}, L., {Snyder}, G.~F., {Satyapal}, S., \& {Ellison}, S.~L. 2018, \mnras, 478, 3056

\bibitem[{{Blumenthal} {et~al.}(2020){Blumenthal}, {Moreno}, {Barnes}, {Hernquist}, {Torrey}, {Claytor}, {Rodriguez-Gomez}, {Marinacci}, \& {Vogelsberger}}]{2020MNRAS.492.2075B}
{Blumenthal}, K.~A., {Moreno}, J., {Barnes}, J.~E., {et~al.} 2020, \mnras, 492, 2075

\bibitem[{{Bottrell} {et~al.}(2019){Bottrell}, {Hani}, {Teimoorinia}, {Ellison}, {Moreno}, {Torrey}, {Hayward}, {Thorp}, {Simard}, \& {Hernquist}}]{2019MNRAS.490.5390B}
{Bottrell}, C., {Hani}, M.~H., {Teimoorinia}, H., {et~al.} 2019, \mnras, 490, 5390

\bibitem[{Bradley {et~al.}(2024)Bradley, Sip{\H o}cz, Robitaille, Tollerud, Vin{\'{\i}}cius, Deil, Barbary, Wilson, Busko, Donath, G{\"u}nther, Cara, Lim, Me{\ss}linger, Burnett, Conseil, Droettboom, Bostroem, Bray, Bratholm, Jamieson, Ginsburg, Barentsen, Craig, Pascual, Rathi, Perrin, Morris, \& Perren}]{larry_bradley_2024_12585239}
Bradley, L., Sip{\H o}cz, B., Robitaille, T., {et~al.} 2024, astropy/photutils: 1.13.0

\bibitem[{{Bruzual} \& {Charlot}(2003)}]{2003MNRAS.344.1000B}
{Bruzual}, G. \& {Charlot}, S. 2003, \mnras, 344, 1000

\bibitem[{{Byrne-Mamahit} {et~al.}(2024){Byrne-Mamahit}, {Patton}, {Ellison}, {Bickley}, {Ferreira}, {Hani}, {Quai}, \& {Wilkinson}}]{2024MNRAS.528.5864B}
{Byrne-Mamahit}, S., {Patton}, D.~R., {Ellison}, S.~L., {et~al.} 2024, \mnras, 528, 5864

\bibitem[{Casey {et~al.}(2023)Casey, Kartaltepe, Drakos, Franco, Harish, Paquereau, Ilbert, Rose, Cox, Nightingale, Robertson, Silverman, Koekemoer, Massey, McCracken, Rhodes, Akins, Amvrosiadis, Arango-Toro, Bagley, Bongiorno, Capak, Champagne, Chartab, Ortiz, Chworowsky, Cooke, Cooper, Darvish, Ding, Faisst, Finkelstein, Fujimoto, Gentile, Gillman, Gould, Gozaliasl, Hayward, He, Hemmati, Hirschmann, Jahnke, Jin, Khostovan, Kokorev, Lambrides, Laigle, Larson, Leung, Liu, Liaudat, Long, Magdis, Mahler, Mainieri, Manning, Maraston, Martin, McCleary, McKinney, McPartland, Mobasher, Pattnaik, Renzini, Rich, Sanders, Sattari, Scognamiglio, Scoville, Sheth, Shuntov, Sparre, Suzuki, Talia, Toft, Trakhtenbrot, Urry, Valentino, Vanderhoof, Vardoulaki, Weaver, Whitaker, Wilkins, Yang, \& Zavala}]{casey2023cosmosweb}
Casey, C.~M., Kartaltepe, J.~S., Drakos, N.~E., {et~al.} 2023, COSMOS-Web: An Overview of the JWST Cosmic Origins Survey

\bibitem[{{{\'C}iprijanovi{\'c}} {et~al.}(2021){{\'C}iprijanovi{\'c}}, {Kafkes}, {Downey}, {Jenkins}, {Perdue}, {Madireddy}, {Johnston}, {Snyder}, \& {Nord}}]{2021MNRAS.506..677C}
{{\'C}iprijanovi{\'c}}, A., {Kafkes}, D., {Downey}, K., {et~al.} 2021, \mnras, 506, 677

\bibitem[{{{\'C}iprijanovi{\'c}} {et~al.}(2020){{\'C}iprijanovi{\'c}}, {Snyder}, {Nord}, \& {Peek}}]{2020A&C....3200390C}
{{\'C}iprijanovi{\'c}}, A., {Snyder}, G.~F., {Nord}, B., \& {Peek}, J.~E.~G. 2020, Astronomy and Computing, 32, 100390

\bibitem[{{Cole} {et~al.}(2000){Cole}, {Lacey}, {Baugh}, \& {Frenk}}]{2000MNRAS.319..168C}
{Cole}, S., {Lacey}, C.~G., {Baugh}, C.~M., \& {Frenk}, C.~S. 2000, \mnras, 319, 168

\bibitem[{{Conselice}(2003)}]{2003ApJS..147....1C}
{Conselice}, C.~J. 2003, \apjs, 147, 1

\bibitem[{{Conselice} {et~al.}(2003){Conselice}, {Bershady}, {Dickinson}, \& {Papovich}}]{2003AJ....126.1183C}
{Conselice}, C.~J., {Bershady}, M.~A., {Dickinson}, M., \& {Papovich}, C. 2003, \aj, 126, 1183

\bibitem[{{Conselice} {et~al.}(2000){Conselice}, {Bershady}, \& {Jangren}}]{2000ApJ...529..886C}
{Conselice}, C.~J., {Bershady}, M.~A., \& {Jangren}, A. 2000, \apj, 529, 886

\bibitem[{{Conselice} {et~al.}(2022){Conselice}, {Mundy}, {Ferreira}, \& {Duncan}}]{2022ApJ...940..168C}
{Conselice}, C.~J., {Mundy}, C.~J., {Ferreira}, L., \& {Duncan}, K. 2022, \apj, 940, 168

\bibitem[{{Darg} {et~al.}(2010){Darg}, {Kaviraj}, {Lintott}, {Schawinski}, {Sarzi}, {Bamford}, {Silk}, {Proctor}, {Andreescu}, {Murray}, {Nichol}, {Raddick}, {Slosar}, {Szalay}, {Thomas}, \& {Vandenberg}}]{2010MNRAS.401.1043D}
{Darg}, D.~W., {Kaviraj}, S., {Lintott}, C.~J., {et~al.} 2010, \mnras, 401, 1043

\bibitem[{Deng {et~al.}(2009)Deng, Dong, Socher, Li, Li, \& Fei-Fei}]{5206848}
Deng, J., Dong, W., Socher, R., {et~al.} 2009, in 2009 IEEE Conference on Computer Vision and Pattern Recognition, 248--255

\bibitem[{{Dieleman} {et~al.}(2015){Dieleman}, {Willett}, \& {Dambre}}]{2015MNRAS.450.1441D}
{Dieleman}, S., {Willett}, K.~W., \& {Dambre}, J. 2015, \mnras, 450, 1441

\bibitem[{{Donnari} {et~al.}(2019){Donnari}, {Pillepich}, {Nelson}, {Vogelsberger}, {Genel}, {Weinberger}, {Marinacci}, {Springel}, \& {Hernquist}}]{2019MNRAS.485.4817D}
{Donnari}, M., {Pillepich}, A., {Nelson}, D., {et~al.} 2019, \mnras, 485, 4817

\bibitem[{{Duan} {et~al.}(2024){Duan}, {Conselice}, {Li}, {Austin}, {Harvey}, {Adams}, {Duncan}, {Trussler}, {Ferreira}, {Westcott}, {Harris}, {Windhorst}, {Holwerda}, {Broadhurst}, {Coe}, {Cohen}, {Driver}, {Frye}, {Grogin}, {Hathi}, {Jansen}, {Koekemoer}, {Marshall}, {Nonino}, {Ortiz}, {Pirzkal}, {Robotham}, {Ryan}, {Summers}, {D'Silva}, {Willmer}, \& {Yan}}]{2024arXiv240709472D}
{Duan}, Q., {Conselice}, C.~J., {Li}, Q., {et~al.} 2024, arXiv e-prints, arXiv:2407.09472

\bibitem[{{Dubois} {et~al.}(2014){Dubois}, {Pichon}, {Welker}, {Le Borgne}, {Devriendt}, {Laigle}, {Codis}, {Pogosyan}, {Arnouts}, {Benabed}, {Bertin}, {Blaizot}, {Bouchet}, {Cardoso}, {Colombi}, {de Lapparent}, {Desjacques}, {Gavazzi}, {Kassin}, {Kimm}, {McCracken}, {Milliard}, {Peirani}, {Prunet}, {Rouberol}, {Silk}, {Slyz}, {Sousbie}, {Teyssier}, {Tresse}, {Treyer}, {Vibert}, \& {Volonteri}}]{2014MNRAS.444.1453D}
{Dubois}, Y., {Pichon}, C., {Welker}, C., {et~al.} 2014, \mnras, 444, 1453

\bibitem[{{Duncan} {et~al.}(2019){Duncan}, {Conselice}, {Mundy}, {Bell}, {Donley}, {Galametz}, {Guo}, {Grogin}, {Hathi}, {Kartaltepe}, {Kocevski}, {Koekemoer}, {P{\'e}rez-Gonz{\'a}lez}, {Mantha}, {Snyder}, \& {Stefanon}}]{2019ApJ...876..110D}
{Duncan}, K., {Conselice}, C.~J., {Mundy}, C., {et~al.} 2019, \apj, 876, 110

\bibitem[{{Ellison} {et~al.}(2008){Ellison}, {Patton}, {Simard}, \& {McConnachie}}]{2008AJ....135.1877E}
{Ellison}, S.~L., {Patton}, D.~R., {Simard}, L., \& {McConnachie}, A.~W. 2008, \aj, 135, 1877

\bibitem[{{Ellison} {et~al.}(2019){Ellison}, {Viswanathan}, {Patton}, {Bottrell}, {McConnachie}, {Gwyn}, \& {Cuillandre}}]{2019MNRAS.487.2491E}
{Ellison}, S.~L., {Viswanathan}, A., {Patton}, D.~R., {et~al.} 2019, \mnras, 487, 2491

\bibitem[{{Ferreira} {et~al.}(2024){Ferreira}, {Bickley}, {Ellison}, {Patton}, {Byrne-Mamahit}, {Wilkinson}, {Bottrell}, {Fabbro}, {Gwyn}, \& {McConnachie}}]{2024MNRAS.533.2547F}
{Ferreira}, L., {Bickley}, R.~W., {Ellison}, S.~L., {et~al.} 2024, \mnras, 533, 2547

\bibitem[{{Ferreira} {et~al.}(2020){Ferreira}, {Conselice}, {Duncan}, {Cheng}, {Griffiths}, \& {Whitney}}]{2020ApJ...895..115F}
{Ferreira}, L., {Conselice}, C.~J., {Duncan}, K., {et~al.} 2020, \apj, 895, 115

\bibitem[{{Gao} {et~al.}(2020){Gao}, {Wang}, {Pearson}, {Gordon}, {Holwerda}, {Hopkins}, {Brown}, {Bland-Hawthorn}, \& {Owers}}]{2020A&A...637A..94G}
{Gao}, F., {Wang}, L., {Pearson}, W.~J., {et~al.} 2020, \aap, 637, A94

\bibitem[{{Gardner} {et~al.}(2006){Gardner}, {Mather}, {Clampin}, {Doyon}, {Greenhouse}, {Hammel}, {Hutchings}, {Jakobsen}, {Lilly}, {Long}, {Lunine}, {McCaughrean}, {Mountain}, {Nella}, {Rieke}, {Rieke}, {Rix}, {Smith}, {Sonneborn}, {Stiavelli}, {Stockman}, {Windhorst}, \& {Wright}}]{2006SSRv..123..485G}
{Gardner}, J.~P., {Mather}, J.~C., {Clampin}, M., {et~al.} 2006, \ssr, 123, 485

\bibitem[{Goodfellow {et~al.}(2016)Goodfellow, Bengio, \& Courville}]{Goodfellow-et-al-2016}
Goodfellow, I., Bengio, Y., \& Courville, A. 2016, Deep Learning (MIT Press), \url{http://www.deeplearningbook.org}

\bibitem[{{Goulding} {et~al.}(2018){Goulding}, {Greene}, {Bezanson}, {Greco}, {Johnson}, {Leauthaud}, {Matsuoka}, {Medezinski}, \& {Price-Whelan}}]{2018PASJ...70S..37G}
{Goulding}, A.~D., {Greene}, J.~E., {Bezanson}, R., {et~al.} 2018, \pasj, 70, S37

\bibitem[{{Helmi}(2020)}]{2020ARA&A..58..205H}
{Helmi}, A. 2020, \araa, 58, 205

\bibitem[{{Hopkins} {et~al.}(2008){Hopkins}, {Hernquist}, {Cox}, \& {Kere{\v{s}}}}]{2008ApJS..175..356H}
{Hopkins}, P.~F., {Hernquist}, L., {Cox}, T.~J., \& {Kere{\v{s}}}, D. 2008, \apjs, 175, 356

\bibitem[{{Huertas-Company} {et~al.}(2015){Huertas-Company}, {Gravet}, {Cabrera-Vives}, {P{\'e}rez-Gonz{\'a}lez}, {Kartaltepe}, {Barro}, {Bernardi}, {Mei}, {Shankar}, {Dimauro}, {Bell}, {Kocevski}, {Koo}, {Faber}, \& {Mcintosh}}]{2015ApJS..221....8H}
{Huertas-Company}, M., {Gravet}, R., {Cabrera-Vives}, G., {et~al.} 2015, \apjs, 221, 8

\bibitem[{{Huertas-Company} \& {Lanusse}(2023)}]{2023PASA...40....1H}
{Huertas-Company}, M. \& {Lanusse}, F. 2023, \pasa, 40, e001

\bibitem[{{Kartaltepe} {et~al.}(2015){Kartaltepe}, {Mozena}, {Kocevski}, {McIntosh}, {Lotz}, {Bell}, {Faber}, {Ferguson}, {Koo}, {Bassett}, {Bernyk}, {Blancato}, {Bournaud}, {Cassata}, {Castellano}, {Cheung}, {Conselice}, {Croton}, {Dahlen}, {de Mello}, {DeGroot}, {Donley}, {Guedes}, {Grogin}, {Hathi}, {Hilton}, {Hollon}, {Koekemoer}, {Liu}, {Lucas}, {Martig}, {McGrath}, {McPartland}, {Mobasher}, {Morlock}, {O'Leary}, {Peth}, {Pforr}, {Pillepich}, {Rosario}, {Soto}, {Straughn}, {Telford}, {Sunnquist}, {Trump}, {Weiner}, {Wuyts}, {Inami}, {Kassin}, {Lani}, {Poole}, \& {Rizer}}]{2015ApJS..221...11K}
{Kartaltepe}, J.~S., {Mozena}, M., {Kocevski}, D., {et~al.} 2015, \apjs, 221, 11

\bibitem[{{Khalid} {et~al.}(2024){Khalid}, {Brough}, {Martin}, {Kimmig}, {Lagos}, {Remus}, \& {Martinez-Lombilla}}]{2024MNRAS.530.4422K}
{Khalid}, A., {Brough}, S., {Martin}, G., {et~al.} 2024, \mnras, 530, 4422

\bibitem[{{Knapen} {et~al.}(2015){Knapen}, {Cisternas}, \& {Querejeta}}]{2015MNRAS.454.1742K}
{Knapen}, J.~H., {Cisternas}, M., \& {Querejeta}, M. 2015, \mnras, 454, 1742

\bibitem[{Krizhevsky {et~al.}(2012)Krizhevsky, Sutskever, \& Hinton}]{article}
Krizhevsky, A., Sutskever, I., \& Hinton, G. 2012, Neural Information Processing Systems, 25

\bibitem[{{La Marca} {et~al.}(2024){La Marca}, {Margalef-Bentabol}, {Wang}, {Gao}, {Goulding}, {Martin}, {Rodriguez-Gomez}, {Trager}, {Yang}, {Dav{\'e}}, \& {Dubois}}]{2024arXiv240718238L}
{La Marca}, A., {Margalef-Bentabol}, B., {Wang}, L., {et~al.} 2024, arXiv e-prints, arXiv:2407.18238

\bibitem[{La~Marca {et~al.}(2024)La~Marca, {Margalef-Bentabol}, Wang, Gao, Goulding, Martin, {Rodriguez-Gomez}, Trager, Yang, Dav{\'e}, \& Dubois}]{lamarcaDustPowerUnravelling2024}
La~Marca, A., {Margalef-Bentabol}, B., Wang, L., {et~al.} 2024, \aap, 690, A326

\bibitem[{{Lambas} {et~al.}(2012){Lambas}, {Alonso}, {Mesa}, \& {O'Mill}}]{2012A&A...539A..45L}
{Lambas}, D.~G., {Alonso}, S., {Mesa}, V., \& {O'Mill}, A.~L. 2012, \aap, 539, A45

\bibitem[{Lecun {et~al.}(1998)Lecun, Bottou, Bengio, \& Haffner}]{726791}
Lecun, Y., Bottou, L., Bengio, Y., \& Haffner, P. 1998, Proceedings of the IEEE, 86, 2278

\bibitem[{{Li} {et~al.}(2020){Li}, {Napolitano}, {Tortora}, {Spiniello}, {Koopmans}, {Huang}, {Roy}, {Vernardos}, {Chatterjee}, {Giblin}, {Getman}, {Radovich}, {Covone}, \& {Kuijken}}]{2020ApJ...899...30L}
{Li}, R., {Napolitano}, N.~R., {Tortora}, C., {et~al.} 2020, \apj, 899, 30

\bibitem[{{Liao} {et~al.}(2023){Liao}, {Johansson}, {Mannerkoski}, {Irodotou}, {Rizzuto}, {McAlpine}, {Rantala}, {Rawlings}, \& {Sawala}}]{2023MNRAS.520.4463L}
{Liao}, S., {Johansson}, P.~H., {Mannerkoski}, M., {et~al.} 2023, \mnras, 520, 4463

\bibitem[{{Lintott} {et~al.}(2011){Lintott}, {Schawinski}, {Bamford}, {Slosar}, {Land}, {Thomas}, {Edmondson}, {Masters}, {Nichol}, {Raddick}, {Szalay}, {Andreescu}, {Murray}, \& {Vandenberg}}]{2011MNRAS.410..166L}
{Lintott}, C., {Schawinski}, K., {Bamford}, S., {et~al.} 2011, \mnras, 410, 166

\bibitem[{{Lintott} {et~al.}(2008){Lintott}, {Schawinski}, {Slosar}, {Land}, {Bamford}, {Thomas}, {Raddick}, {Nichol}, {Szalay}, {Andreescu}, {Murray}, \& {Vandenberg}}]{2008MNRAS.389.1179L}
{Lintott}, C.~J., {Schawinski}, K., {Slosar}, A., {et~al.} 2008, \mnras, 389, 1179

\bibitem[{{Lotz} {et~al.}(2008){Lotz}, {Davis}, {Faber}, {Guhathakurta}, {Gwyn}, {Huang}, {Koo}, {Le Floc'h}, {Lin}, {Newman}, {Noeske}, {Papovich}, {Willmer}, {Coil}, {Conselice}, {Cooper}, {Hopkins}, {Metevier}, {Primack}, {Rieke}, \& {Weiner}}]{2008ApJ...672..177L}
{Lotz}, J.~M., {Davis}, M., {Faber}, S.~M., {et~al.} 2008, \apj, 672, 177

\bibitem[{{Lotz} {et~al.}(2011){Lotz}, {Jonsson}, {Cox}, {Croton}, {Primack}, {Somerville}, \& {Stewart}}]{2011ApJ...742..103L}
{Lotz}, J.~M., {Jonsson}, P., {Cox}, T.~J., {et~al.} 2011, \apj, 742, 103

\bibitem[{{Lotz} {et~al.}(2004){Lotz}, {Primack}, \& {Madau}}]{2004AJ....128..163L}
{Lotz}, J.~M., {Primack}, J., \& {Madau}, P. 2004, \aj, 128, 163

\bibitem[{Lupton {et~al.}(2004)Lupton, Blanton, Fekete, Hogg, O’Mullane, Szalay, \& Wherry}]{Lupton_2004}
Lupton, R., Blanton, M., Fekete, G., {et~al.} 2004, Publications of the Astronomical Society of the Pacific, 116, 133–137

\bibitem[{{Margalef-Bentabol} {et~al.}(2024){Margalef-Bentabol}, {Wang}, {La Marca}, {Blanco-Prieto}, {Chudy}, {Dom{\'\i}nguez-S{\'a}nchez}, {Goulding}, {Guzm{\'a}n-Ortega}, {Huertas-Company}, {Martin}, {Pearson}, {Rodriguez-Gomez}, {Walmsley}, {Bickley}, {Bottrell}, {Conselice}, \& {O'Ryan}}]{2024A&A...687A..24M}
{Margalef-Bentabol}, B., {Wang}, L., {La Marca}, A., {et~al.} 2024, \aap, 687, A24

\bibitem[{{Marinacci} {et~al.}(2018){Marinacci}, {Vogelsberger}, {Pakmor}, {Torrey}, {Springel}, {Hernquist}, {Nelson}, {Weinberger}, {Pillepich}, {Naiman}, \& {Genel}}]{2018MNRAS.480.5113M}
{Marinacci}, F., {Vogelsberger}, M., {Pakmor}, R., {et~al.} 2018, \mnras, 480, 5113

\bibitem[{{Naiman} {et~al.}(2018){Naiman}, {Pillepich}, {Springel}, {Ramirez-Ruiz}, {Torrey}, {Vogelsberger}, {Pakmor}, {Nelson}, {Marinacci}, {Hernquist}, {Weinberger}, \& {Genel}}]{2018MNRAS.477.1206N}
{Naiman}, J.~P., {Pillepich}, A., {Springel}, V., {et~al.} 2018, \mnras, 477, 1206

\bibitem[{Nelson {et~al.}(2017)Nelson, Pillepich, Springel, Weinberger, Hernquist, Pakmor, Genel, Torrey, Vogelsberger, Kauffmann, Marinacci, \& Naiman}]{Nelson_2017}
Nelson, D., Pillepich, A., Springel, V., {et~al.} 2017, Monthly Notices of the Royal Astronomical Society, 475, 624–647

\bibitem[{{Nelson} {et~al.}(2019){Nelson}, {Springel}, {Pillepich}, {Rodriguez-Gomez}, {Torrey}, {Genel}, {Vogelsberger}, {Pakmor}, {Marinacci}, {Weinberger}, {Kelley}, {Lovell}, {Diemer}, \& {Hernquist}}]{2019ComAC...6....2N}
{Nelson}, D., {Springel}, V., {Pillepich}, A., {et~al.} 2019, Computational Astrophysics and Cosmology, 6, 2

\bibitem[{Nelson {et~al.}(2021)Nelson, Springel, Pillepich, Rodriguez-Gomez, Torrey, Genel, Vogelsberger, Pakmor, Marinacci, Weinberger, Kelley, Lovell, Diemer, \& Hernquist}]{nelson2021illustristng}
Nelson, D., Springel, V., Pillepich, A., {et~al.} 2021, The IllustrisTNG Simulations: Public Data Release

\bibitem[{{Patton} {et~al.}(2011){Patton}, {Ellison}, {Simard}, {McConnachie}, \& {Mendel}}]{2011MNRAS.412..591P}
{Patton}, D.~R., {Ellison}, S.~L., {Simard}, L., {McConnachie}, A.~W., \& {Mendel}, J.~T. 2011, \mnras, 412, 591

\bibitem[{{Pearson} {et~al.}(2024){Pearson}, {Rodriguez-Gomez}, {Kruk}, \& {Margalef-Bentabol}}]{2024A&A...687A..45P}
{Pearson}, W.~J., {Rodriguez-Gomez}, V., {Kruk}, S., \& {Margalef-Bentabol}, B. 2024, \aap, 687, A45

\bibitem[{{Pearson} {et~al.}(2022){Pearson}, {Suelves}, {Ho}, {Oi}, {Brough}, {Holwerda}, {Hopkins}, {Huang}, {Hwang}, {Kelvin}, {Kim}, {L{\'o}pez-S{\'a}nchez}, {Ma{\l}ek}, {Pearson}, {Poliszczuk}, {Pollo}, {Rodriguez-Gomez}, {Shim}, {Toba}, \& {Wang}}]{2022A&A...661A..52P}
{Pearson}, W.~J., {Suelves}, L.~E., {Ho}, S.~C.~C., {et~al.} 2022, \aap, 661, A52

\bibitem[{{Pearson} {et~al.}(2019{\natexlab{a}}){Pearson}, {Wang}, {Alpaslan}, {Baldry}, {Bilicki}, {Brown}, {Grootes}, {Holwerda}, {Kitching}, {Kruk}, \& {van der Tak}}]{2019A&A...631A..51P}
{Pearson}, W.~J., {Wang}, L., {Alpaslan}, M., {et~al.} 2019{\natexlab{a}}, \aap, 631, A51

\bibitem[{{Pearson} {et~al.}(2019{\natexlab{b}}){Pearson}, {Wang}, {Trayford}, {Petrillo}, \& {van der Tak}}]{2019A&A...626A..49P}
{Pearson}, W.~J., {Wang}, L., {Trayford}, J.~W., {Petrillo}, C.~E., \& {van der Tak}, F.~F.~S. 2019{\natexlab{b}}, \aap, 626, A49

\bibitem[{{Petrillo} {et~al.}(2017){Petrillo}, {Tortora}, {Chatterjee}, {Vernardos}, {Koopmans}, {Verdoes Kleijn}, {Napolitano}, {Covone}, {Schneider}, {Grado}, \& {McFarland}}]{2017MNRAS.472.1129P}
{Petrillo}, C.~E., {Tortora}, C., {Chatterjee}, S., {et~al.} 2017, \mnras, 472, 1129

\bibitem[{Pillepich {et~al.}(2017)Pillepich, Nelson, Hernquist, Springel, Pakmor, Torrey, Weinberger, Genel, Naiman, Marinacci, \& Vogelsberger}]{10.1093/mnras/stx3112}
Pillepich, A., Nelson, D., Hernquist, L., {et~al.} 2017, Monthly Notices of the Royal Astronomical Society, 475, 648

\bibitem[{{Planck Collaboration} {et~al.}(2016){Planck Collaboration}, {Ade}, {Aghanim}, {Arnaud}, {Ashdown}, {Aumont}, {Baccigalupi}, {Banday}, {Barreiro}, {Bartlett}, {Bartolo}, {Battaner}, {Battye}, {Benabed}, {Beno{\^\i}t}, {Benoit-L{\'e}vy}, {Bernard}, {Bersanelli}, {Bielewicz}, {Bock}, {Bonaldi}, {Bonavera}, {Bond}, {Borrill}, {Bouchet}, {Boulanger}, {Bucher}, {Burigana}, {Butler}, {Calabrese}, {Cardoso}, {Catalano}, {Challinor}, {Chamballu}, {Chary}, {Chiang}, {Chluba}, {Christensen}, {Church}, {Clements}, {Colombi}, {Colombo}, {Combet}, {Coulais}, {Crill}, {Curto}, {Cuttaia}, {Danese}, {Davies}, {Davis}, {de Bernardis}, {de Rosa}, {de Zotti}, {Delabrouille}, {D{\'e}sert}, {Di Valentino}, {Dickinson}, {Diego}, {Dolag}, {Dole}, {Donzelli}, {Dor{\'e}}, {Douspis}, {Ducout}, {Dunkley}, {Dupac}, {Efstathiou}, {Elsner}, {En{\ss}lin}, {Eriksen}, {Farhang}, {Fergusson}, {Finelli}, {Forni}, {Frailis}, {Fraisse}, {Franceschi}, {Frejsel}, {Galeotta}, {Galli}, {Ganga}, {Gauthier}, {Gerbino}, {Ghosh}, {Giard},
  {Giraud-H{\'e}raud}, {Giusarma}, {Gjerl{\o}w}, {Gonz{\'a}lez-Nuevo}, {G{\'o}rski}, {Gratton}, {Gregorio}, {Gruppuso}, {Gudmundsson}, {Hamann}, {Hansen}, {Hanson}, {Harrison}, {Helou}, {Henrot-Versill{\'e}}, {Hern{\'a}ndez-Monteagudo}, {Herranz}, {Hildebrandt}, {Hivon}, {Hobson}, {Holmes}, {Hornstrup}, {Hovest}, {Huang}, {Huffenberger}, {Hurier}, {Jaffe}, {Jaffe}, {Jones}, {Juvela}, {Keih{\"a}nen}, {Keskitalo}, {Kisner}, {Kneissl}, {Knoche}, {Knox}, {Kunz}, {Kurki-Suonio}, {Lagache}, {L{\"a}hteenm{\"a}ki}, {Lamarre}, {Lasenby}, {Lattanzi}, {Lawrence}, {Leahy}, {Leonardi}, {Lesgourgues}, {Levrier}, {Lewis}, {Liguori}, {Lilje}, {Linden-V{\o}rnle}, {L{\'o}pez-Caniego}, {Lubin}, {Mac{\'\i}as-P{\'e}rez}, {Maggio}, {Maino}, {Mandolesi}, {Mangilli}, {Marchini}, {Maris}, {Martin}, {Martinelli}, {Mart{\'\i}nez-Gonz{\'a}lez}, {Masi}, {Matarrese}, {McGehee}, {Meinhold}, {Melchiorri}, {Melin}, {Mendes}, {Mennella}, {Migliaccio}, {Millea}, {Mitra}, {Miville-Desch{\^e}nes}, {Moneti}, {Montier}, {Morgante}, {Mortlock},
  {Moss}, {Munshi}, {Murphy}, {Naselsky}, {Nati}, {Natoli}, {Netterfield}, {N{\o}rgaard-Nielsen}, {Noviello}, {Novikov}, {Novikov}, {Oxborrow}, {Paci}, {Pagano}, {Pajot}, {Paladini}, {Paoletti}, {Partridge}, {Pasian}, {Patanchon}, {Pearson}, {Perdereau}, {Perotto}, {Perrotta}, {Pettorino}, {Piacentini}, {Piat}, {Pierpaoli}, {Pietrobon}, {Plaszczynski}, {Pointecouteau}, {Polenta}, {Popa}, {Pratt}, {Pr{\'e}zeau}, {Prunet}, {Puget}, {Rachen}, {Reach}, {Rebolo}, {Reinecke}, {Remazeilles}, {Renault}, {Renzi}, {Ristorcelli}, {Rocha}, {Rosset}, {Rossetti}, {Roudier}, {Rouill{\'e} d'Orfeuil}, {Rowan-Robinson}, {Rubi{\~n}o-Mart{\'\i}n}, {Rusholme}, {Said}, {Salvatelli}, {Salvati}, {Sandri}, {Santos}, {Savelainen}, {Savini}, {Scott}, {Seiffert}, {Serra}, {Shellard}, {Spencer}, {Spinelli}, {Stolyarov}, {Stompor}, {Sudiwala}, {Sunyaev}, {Sutton}, {Suur-Uski}, {Sygnet}, {Tauber}, {Terenzi}, {Toffolatti}, {Tomasi}, {Tristram}, {Trombetti}, {Tucci}, {Tuovinen}, {T{\"u}rler}, {Umana}, {Valenziano}, {Valiviita}, {Van Tent},
  {Vielva}, {Villa}, {Wade}, {Wandelt}, {Wehus}, {White}, {White}, {Wilkinson}, {Yvon}, {Zacchei}, \& {Zonca}}]{2016A&A...594A..13P}
{Planck Collaboration}, {Ade}, P.~A.~R., {Aghanim}, N., {et~al.} 2016, \aap, 594, A13

\bibitem[{{Quai} {et~al.}(2023){Quai}, {Byrne-Mamahit}, {Ellison}, {Patton}, \& {Hani}}]{2023MNRAS.519.2119Q}
{Quai}, S., {Byrne-Mamahit}, S., {Ellison}, S.~L., {Patton}, D.~R., \& {Hani}, M.~H. 2023, \mnras, 519, 2119

\bibitem[{{Rieke} {et~al.}(2023){Rieke}, {Kelly}, {Misselt}, {Stansberry}, {Boyer}, {Beatty}, {Egami}, {Florian}, {Greene}, {Hainline}, {Leisenring}, {Roellig}, {Schlawin}, {Sun}, {Tinnin}, {Williams}, {Willmer}, {Wilson}, {Clark}, {Rohrbach}, {Brooks}, {Canipe}, {Correnti}, {DiFelice}, {Gennaro}, {Girard}, {Hartig}, {Hilbert}, {Koekemoer}, {Nikolov}, {Pirzkal}, {Rest}, {Robberto}, {Sunnquist}, {Telfer}, {Wu}, {Ferry}, {Lewis}, {Baum}, {Beichman}, {Doyon}, {Dressler}, {Eisenstein}, {Ferrarese}, {Hodapp}, {Horner}, {Jaffe}, {Johnstone}, {Krist}, {Martin}, {McCarthy}, {Meyer}, {Rieke}, {Trauger}, \& {Young}}]{2023PASP..135b8001R}
{Rieke}, M.~J., {Kelly}, D.~M., {Misselt}, K., {et~al.} 2023, \pasp, 135, 028001

\bibitem[{Rodriguez-Gomez {et~al.}(2015)Rodriguez-Gomez, Genel, Vogelsberger, Sijacki, Pillepich, Sales, Torrey, Snyder, Nelson, Springel, Ma, \& Hernquist}]{Rodriguez_Gomez_2015}
Rodriguez-Gomez, V., Genel, S., Vogelsberger, M., {et~al.} 2015, Monthly Notices of the Royal Astronomical Society, 449, 49–64

\bibitem[{{Rodr{\'\i}guez Montero} {et~al.}(2019){Rodr{\'\i}guez Montero}, {Dav{\'e}}, {Wild}, {Angl{\'e}s-Alc{\'a}zar}, \& {Narayanan}}]{2019MNRAS.490.2139R}
{Rodr{\'\i}guez Montero}, F., {Dav{\'e}}, R., {Wild}, V., {Angl{\'e}s-Alc{\'a}zar}, D., \& {Narayanan}, D. 2019, \mnras, 490, 2139

\bibitem[{{Satyapal} {et~al.}(2014){Satyapal}, {Ellison}, {McAlpine}, {Hickox}, {Patton}, \& {Mendel}}]{2014MNRAS.441.1297S}
{Satyapal}, S., {Ellison}, S.~L., {McAlpine}, W., {et~al.} 2014, \mnras, 441, 1297

\bibitem[{{Schaefer} {et~al.}(2018){Schaefer}, {Geiger}, {Kuntzer}, \& {Kneib}}]{2018A&A...611A...2S}
{Schaefer}, C., {Geiger}, M., {Kuntzer}, T., \& {Kneib}, J.~P. 2018, \aap, 611, A2

\bibitem[{{Schaye} {et~al.}(2015){Schaye}, {Crain}, {Bower}, {Furlong}, {Schaller}, {Theuns}, {Dalla Vecchia}, {Frenk}, {McCarthy}, {Helly}, {Jenkins}, {Rosas-Guevara}, {White}, {Baes}, {Booth}, {Camps}, {Navarro}, {Qu}, {Rahmati}, {Sawala}, {Thomas}, \& {Trayford}}]{2015MNRAS.446..521S}
{Schaye}, J., {Crain}, R.~A., {Bower}, R.~G., {et~al.} 2015, \mnras, 446, 521

\bibitem[{{Scoville} {et~al.}(2007){Scoville}, {Aussel}, {Brusa}, {Capak}, {Carollo}, {Elvis}, {Giavalisco}, {Guzzo}, {Hasinger}, {Impey}, {Kneib}, {LeFevre}, {Lilly}, {Mobasher}, {Renzini}, {Rich}, {Sanders}, {Schinnerer}, {Schminovich}, {Shopbell}, {Taniguchi}, \& {Tyson}}]{2007ApJS..172....1S}
{Scoville}, N., {Aussel}, H., {Brusa}, M., {et~al.} 2007, \apjs, 172, 1

\bibitem[{{Selvaraju} {et~al.}(2016){Selvaraju}, {Cogswell}, {Das}, {Vedantam}, {Parikh}, \& {Batra}}]{2016arXiv161002391S}
{Selvaraju}, R.~R., {Cogswell}, M., {Das}, A., {et~al.} 2016, arXiv e-prints, arXiv:1610.02391

\bibitem[{{Snyder} {et~al.}(2019){Snyder}, {Rodriguez-Gomez}, {Lotz}, {Torrey}, {Quirk}, {Hernquist}, {Vogelsberger}, \& {Freeman}}]{2019MNRAS.486.3702S}
{Snyder}, G.~F., {Rodriguez-Gomez}, V., {Lotz}, J.~M., {et~al.} 2019, \mnras, 486, 3702

\bibitem[{Springel {et~al.}(2017)Springel, Pakmor, Pillepich, Weinberger, Nelson, Hernquist, Vogelsberger, Genel, Torrey, Marinacci, \& Naiman}]{10.1093/mnras/stx3304}
Springel, V., Pakmor, R., Pillepich, A., {et~al.} 2017, Monthly Notices of the Royal Astronomical Society, 475, 676

\bibitem[{Tan \& Le(2020)}]{tan2020efficientnet}
Tan, M. \& Le, Q.~V. 2020, EfficientNet: Rethinking Model Scaling for Convolutional Neural Networks

\bibitem[{{Villforth}(2023)}]{2023OJAp....6E..34V}
{Villforth}, C. 2023, The Open Journal of Astrophysics, 6, 34

\bibitem[{{Vogelsberger} {et~al.}(2014){Vogelsberger}, {Genel}, {Springel}, {Torrey}, {Sijacki}, {Xu}, {Snyder}, {Nelson}, \& {Hernquist}}]{2014MNRAS.444.1518V}
{Vogelsberger}, M., {Genel}, S., {Springel}, V., {et~al.} 2014, \mnras, 444, 1518

\bibitem[{{Walmsley} {et~al.}(2023){Walmsley}, {Allen}, {Aussel}, {Bowles}, {Gregorowicz}, {Slijepcevic}, {Lintott}, {Scaife}, {Jab{\l}o{\'n}ska}, {Karchev}, {Lanzieri}, {Mohan}, {O'Ryan}, {Saiguhan}, {Su{\'a}rez}, {Guerra-Varas}, \& {Velu}}]{2023JOSS....8.5312W}
{Walmsley}, M., {Allen}, C., {Aussel}, B., {et~al.} 2023, The Journal of Open Source Software, 8, 5312

\bibitem[{Walmsley {et~al.}(2023)Walmsley, Allen, Aussel, Bowles, Gregorowicz, Slijepcevic, Lintott, m.~Scaife, Jabłońska, Karchev, Lanzieri, Mohan, O’Ryan, Saiguhan, Suárez, Guerra-Varas, \& Velu}]{Walmsley2023}
Walmsley, M., Allen, C., Aussel, B., {et~al.} 2023, Journal of Open Source Software, 8, 5312

\bibitem[{{Walmsley} {et~al.}(2022){Walmsley}, {Lintott}, {G{\'e}ron}, {Kruk}, {Krawczyk}, {Willett}, {Bamford}, {Kelvin}, {Fortson}, {Gal}, {Keel}, {Masters}, {Mehta}, {Simmons}, {Smethurst}, {Smith}, {Baeten}, \& {Macmillan}}]{2022MNRAS.509.3966W}
{Walmsley}, M., {Lintott}, C., {G{\'e}ron}, T., {et~al.} 2022, \mnras, 509, 3966

\bibitem[{Walmsley {et~al.}(2021)Walmsley, Lintott, Géron, Kruk, Krawczyk, Willett, Bamford, Kelvin, Fortson, Gal, Keel, Masters, Mehta, Simmons, Smethurst, Smith, Baeten, \& Macmillan}]{10.1093/mnras/stab2093}
Walmsley, M., Lintott, C., Géron, T., {et~al.} 2021, Monthly Notices of the Royal Astronomical Society, 509, 3966

\bibitem[{{Wang} {et~al.}(2020){Wang}, {Pearson}, \& {Rodriguez-Gomez}}]{2020A&A...644A..87W}
{Wang}, L., {Pearson}, W.~J., \& {Rodriguez-Gomez}, V. 2020, \aap, 644, A87

\bibitem[{{Weaver} {et~al.}(2022){Weaver}, {Kauffmann}, {Ilbert}, {McCracken}, {Moneti}, {Toft}, {Brammer}, {Shuntov}, {Davidzon}, {Hsieh}, {Laigle}, {Anastasiou}, {Jespersen}, {Vinther}, {Capak}, {Casey}, {McPartland}, {Milvang-Jensen}, {Mobasher}, {Sanders}, {Zalesky}, {Arnouts}, {Aussel}, {Dunlop}, {Faisst}, {Franx}, {Furtak}, {Fynbo}, {Gould}, {Greve}, {Gwyn}, {Kartaltepe}, {Kashino}, {Koekemoer}, {Kokorev}, {Le F{\`e}vre}, {Lilly}, {Masters}, {Magdis}, {Mehta}, {Peng}, {Riechers}, {Salvato}, {Sawicki}, {Scarlata}, {Scoville}, {Shirley}, {Silverman}, {Sneppen}, {Smolc̆i{\'c}}, {Steinhardt}, {Stern}, {Tanaka}, {Taniguchi}, {Teplitz}, {Vaccari}, {Wang}, \& {Zamorani}}]{2022ApJS..258...11W}
{Weaver}, J.~R., {Kauffmann}, O.~B., {Ilbert}, O., {et~al.} 2022, \apjs, 258, 11

\bibitem[{{White} \& {Frenk}(1991)}]{1991ApJ...379...52W}
{White}, S. D.~M. \& {Frenk}, C.~S. 1991, \apj, 379, 52

\bibitem[{{Zanisi} {et~al.}(2021){Zanisi}, {Huertas-Company}, {Lanusse}, {Bottrell}, {Pillepich}, {Nelson}, {Rodriguez-Gomez}, {Shankar}, {Hernquist}, {Dekel}, {Margalef-Bentabol}, {Vogelsberger}, \& {Primack}}]{2021MNRAS.501.4359Z}
{Zanisi}, L., {Huertas-Company}, M., {Lanusse}, F., {et~al.} 2021, \mnras, 501, 4359

\bibitem[{Zhuang {et~al.}(2023)Zhuang, Li, \& Shen}]{zhuang2023agns}
Zhuang, M.-Y., Li, J., \& Shen, Y. 2023, AGNs and Host Galaxies in COSMOS-Web. I. NIRCam Images, PSF Models and Initial Results on X-ray-selected Broad-line AGNs at $0.35\lesssim z \lesssim 3.5$

\end{thebibliography}

\begin{appendix}

\section{Hyper-parameter tuning}
\label{appendix:tuning}

The final grid search was performed over batch size, learning rate, weight decay, and the number of blocks that the optimiser will consider to update. For each grid search, only a subset of the complete catalogue at low redshifts ($z=0.5$) was considered to reduce computation time. In total, we did three grid searches: one to train the model of the one-stage classification set-up, and two to train the  models used for the two-stage classification set-up. The parameters considered for each grid search can be found in Table \ref{table: sweep}. 
The grid search was created and performed via the Weights \& Biases (WandB) platform \citep{wandb}. This artificial intelligence (AI) developer platform not only provides an interface to create and initiate the grid search on, but also tracks the performance metrics for each run. In the WandB interface, the performance metrics are plotted for both the validation and training data as a function of epoch. 
We choose to present in this paper only the accuracy performance metric, but recall and precision were also considered during the decision making process on the hyperparameter search.

The grid search for the first model of the two-stage set-up (i.e. the merger/non-merger classification), resulted in 44 runs. It is visually difficult to distinguish each run and conclude which configuration is best. Fortunately, WandB provides multiple solutions that help identifying the effect that each parameter has. Instead of looking at all configurations simultaneously, it is also possible to group the runs on a specific parameter. 
For instance, if we group by batch size, which in our case has two values, then the results are two plots with one showing the average performance over all runs consisting of the first value and the other showing the average performance over all runs consisting of the other value. From this average behaviour, we can identify the influence that a given parameter has on the learning curve. 
Another useful tool provided by WandB is the filter. Once we have established that a particular value works best for a specific hyperparameter, we are not interested anymore in the runs consisting of a different value. Using WandB's filter function, we can extract only those runs we are interested in. 


\begin{table*}
    \centering
    \caption{Hyperparameter settings for each grid search.}
    \begin{tabular}{|c|c|c|c|c|c||}
    \hline
    Classification set-up & Batch size & Number of blocks & Learning rate & Weight decay \\
    \hline 
      One-stage & 128, 256 & 2, 4, 7 & 1e-4, 1e-5 & 0.05, 0.005\\
      \hline
    Two-stage step 1 & 64, 128  & 2, 7 & 1e-4, 1e-5 & 0.5, 0.05 \\
     \hline
    Two-stage step 2 & 32, 128 & 2, 4, 7 & 1e-4, 1e-5 & 0.5, 0.05, 0.005\\
    \hline
    \end{tabular}
     \tablefoot{The final grid search is performed over the batch size, learning rate, weight decay, and the number of blocks that the optimiser will consider to update. For each grid search, only a subset of the complete catalogue at low redshifts is considered. We have a total of  three grid searches, one for the model in the one-stage classification set-up and two for the models used in the two-stage classification set-up.}
    \label{table: sweep}
\end{table*}

\begin{figure*}
\centering
\includegraphics[width=0.45\textwidth]{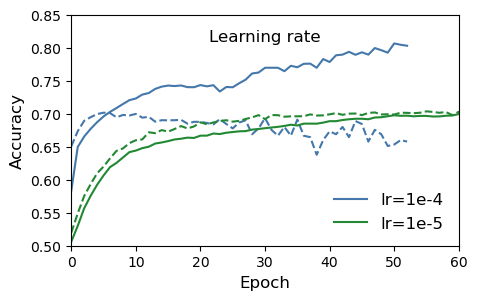} 
\includegraphics[width=0.45\textwidth]{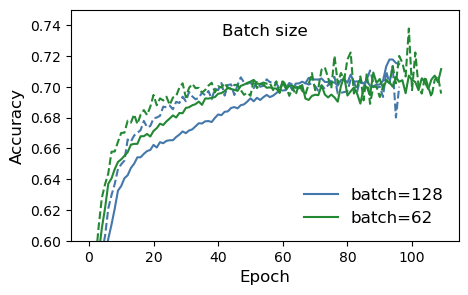}
\includegraphics[width=0.45\textwidth]{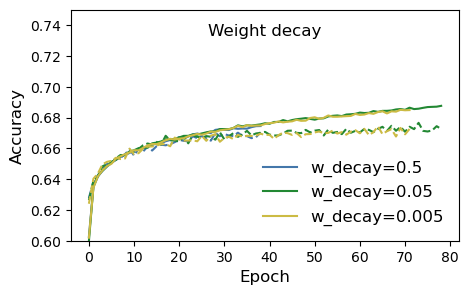}
\includegraphics[width=0.45\textwidth]{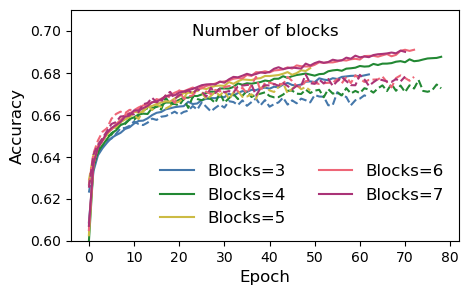}
\caption{Accuracy scores of the merger/non-merger classification versus epoch during the training phase. The top two graphs are the result of the grid search performed on  galaxies at low redshifts. The bottom two graphs are the results of training on the entire dataset ranging over redshift $z=0$ to 3. The title of each panel indicates the parameter which the runs are grouped on. In other words, each panel shows the average scores of all runs containing the parameter values shown in the legend.  
(a) Grid search grouped on the learning rate parameter.
(b) Grid search grouped on the batch size parameter.
(c) Separate fine-tuning runs with different weight decay values. 
(d) Separate fine-tuning runs by tuning deeper into the blocks. The solid lines correspond to the training set and the dashed lines correspond to the validation set.}
\label{fig:gridsearch}
\end{figure*}

\begin{figure}
    \centering
    \includegraphics[width=0.85\linewidth]{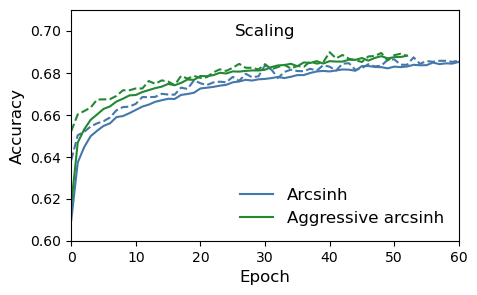}
    \caption{Separate runs performed on the scaling while leaving the other hyperparameters the same. The blue lines correspond to the simple arcsinh scaling, while the green lines used a more aggressive arcsinh scaling which decreases the presence of the background further. The solid lines correspond to the training set and the dashed lines correspond to the validation set.}
    \label{fig:param search scaling}
\end{figure}

For the merger/non-merger classification we first grouped the runs based on learning rate. The top left panel in Fig. \ref{fig:gridsearch} shows the accuracy of the training data and validation data for  runs with learning rate of $1e-4$ and $1e-5$. At first, it seem that the higher learning rate performs better. However, when focusing only on the validation data, one can see that they reach the same accuracy. This indicates that a higher learning rate will tend to overfit. Therefore, we filtered out all runs with higher learning rates. The same trend is seen in the pre-and post-merger classification model, which we do not visualise here.
Next we grouped the runs on batch size, as shown in the top right panel in  Fig. \ref{fig:gridsearch}. For the validation dataset, runs with different batch sizes reach similar accuracies. For the training dataset, the batch size of $64$ is slightly better. As we want to keep performance for the training and validation datasets as close as possible to reduce overfitting, a batch size of $128$ is better. Note that for the batch size we discussed the merger/non-merger classification only. The pre-merger/post-merger classification did not show any clear preference. Since for the two-stage set-up we combine the two models of each stage, we decided to keep the parameters for both stages the same.

For the last two parameters we did separate runs on the complete dataset over the redshift range $0.5<z<3$ as the grid search did not provide any clear distinction. Changing the weight decay parameter from the default ($0.05$) had no influence on the training. The most apparent impact on the performance is the number of epochs it takes to reach the stopping criteria. Increasing the weight decay lead to an early-stopping. However, it would need more epochs during training to reach the accuracy of the other runs with lower weight decay, as demonstrated in the lower left panel in Fig. \ref{fig:gridsearch}. Therefore, we decided to use the default value.
We also performed various runs in which, with each run, we fine-tuned one deeper block. 
Fine-tuning up to block $3$, $4$, $5$, $6$ and $7$ is shown in the lower right panel in Fig. \ref{fig:gridsearch}. The highest accuracy is achieved when increasing the number of blocks to perform the fine-tuning on. However, finetuning on more blocks increases computation time. Therefore, we chose to finetune 4 blocks in the first stage model, and 5 blocks in the second model. 
Lastly, we also performed two runs for the two scaling methods as discussed in Sect. \ref{sect: mock}. 
The resulting two runs are shown in Fig. \ref{fig:param search scaling}. We conclude that the aggressive arcsinh scaling leads to higher accuracies.
The plots resulted only from the grid search and some consecutive runs done on the first model in the two-stage classification. For the results of the other two models we have used the same procedure. In general, the same trends among the parameters were seen in all runs.

\section{Binary classification: merger vs. non-merger}
\label{appendix:binary}

In \citet{2024A&A...687A..24M}, six leading machine learning/deep learning based merger detection methods were bench-marked using the same datasets, which include mock observations from the IllustrisTNG simulations, mock observations from the Horizon-AGN simulations and real observations from the Hyper Suprime-Cam Subaru Strategic Program (HSC-SSP) survey. In the redshift range $0.52 \leq z\leq 0.76$, the precision for the merger class achieved by these methods  on the TNG-test dataset ranges between 68.1\% to 78.6\% and the recall for the merger class ranges between 66.4\% and 84.1\%. In the redshift range $0.76 \leq z\leq 1$, precision ranges between 68.2\% to 75.7\% and recall ranges between 57.5\% to 76.2\%.

\begin{figure*}
    \centering
    \includegraphics[width=0.35\linewidth]{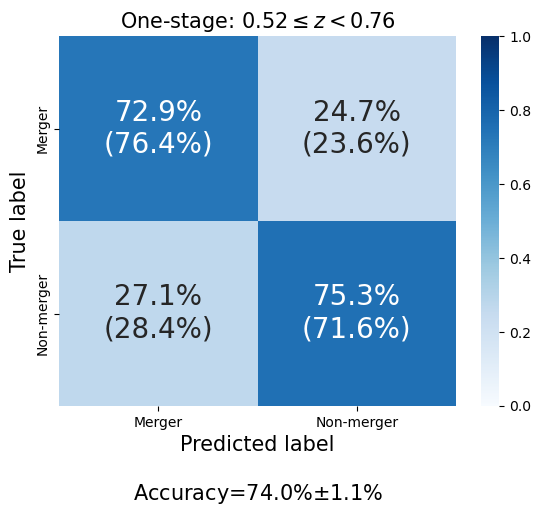}
    \includegraphics[width=0.35\linewidth]{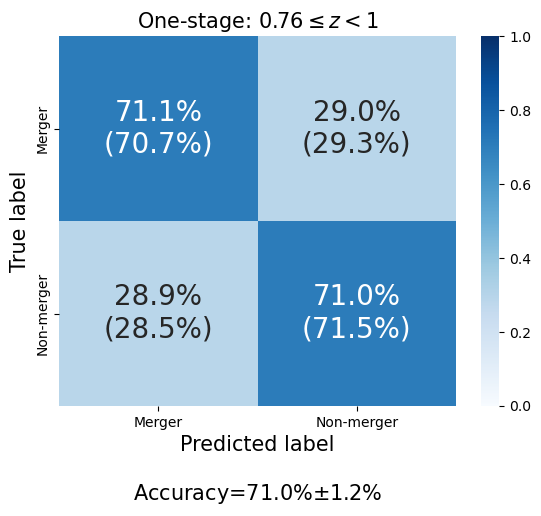}
    \includegraphics[width=0.35\linewidth]{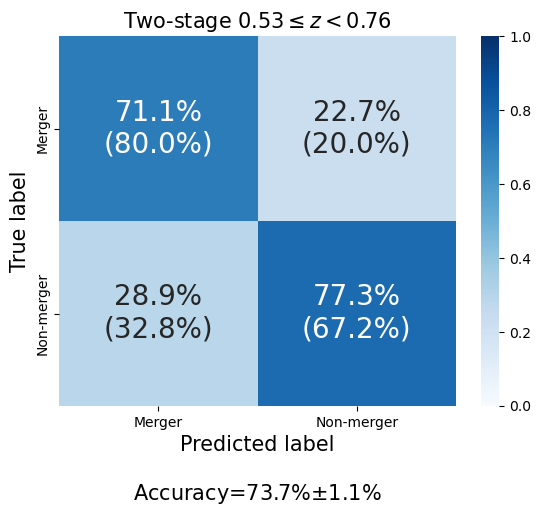}
    \includegraphics[width=0.35\linewidth]{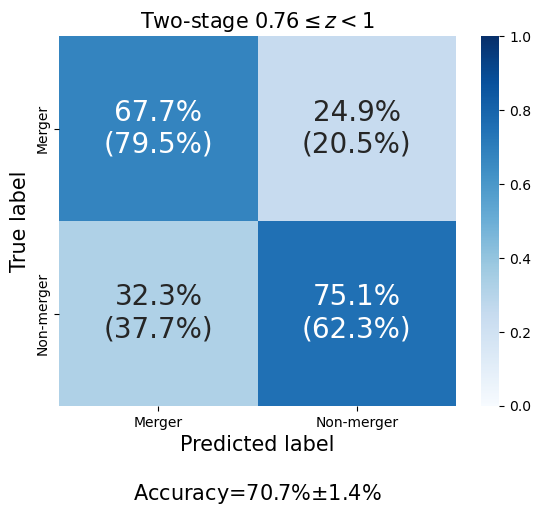}
    \caption{Top left: Confusion matrix for the one-stage classification in the binary classification case (i.e., mergers vs non-mergers), predicted on the data set in the redshift range $0.52 \leq z\leq 0.76$. Top right: Similar to the top left figure but for $0.76 \leq z\leq 1$. Bottom left: Confusion matrix for the two-stage classification in the binary classification case, in the redshift range $0.52 \leq z\leq 0.76$. Bottom right: Similar to the bottom left figure but for $0.76 \leq z\leq 1$.}
    \label{fig:binary_classification}
\end{figure*}

To directly compare with the performances achieved in \citet{2024A&A...687A..24M}, we combined our pre- and post-merger classes into a single merger class and re-calculated the performance metrics of our classifiers in the binary classification case (i.e., mergers vs. non-mergers). In Fig. \ref{fig:binary_classification}, we show the confusion matrices for binary classification from our one-stage and two-stage classification set-ups. It is clear that our classifiers presented in this paper achieve similar performance to the methods presented in \citet{2024A&A...687A..24M}.

\end{appendix}


\end{document}